\documentclass[twocolumn,superscriptaddress,amsmath,amssymb,aps]{revtex4-2}
\usepackage{graphicx}
\usepackage{float}
\usepackage{color}
\usepackage{bbm}
\usepackage{multirow}
\usepackage{subfigure}
\usepackage{calc}
\usepackage{bm}
\usepackage{epstopdf}
\usepackage{mathrsfs}
\usepackage[colorlinks=true,linkcolor=blue,urlcolor=blue,citecolor=blue,anchorcolor=blue]{hyperref}
\usepackage{booktabs}
\usepackage{array}
\usepackage[table]{xcolor}
\usepackage{comment}

\usepackage{xr}
\usepackage{siunitx}
\DeclareSIUnit{\um}{\micro\meter}

\begin{document}

\title{Universal scaling laws for dynamical-thermal hysteresis}
 
\author{Yachao Sun}
\thanks{These two authors contributed equally}
\affiliation{Institute of Theoretical Physics, Chinese Academy of Sciences, Beijing 100190, China}

\author{Xuesong Li}
\thanks{These two authors contributed equally}
\affiliation{Dongguan Institute of Materials Science and Technology, Chinese Academy of Sciences, Dongguan 523808, China}
\affiliation{Songshan Lake Materials Laboratory, Dongguan 523808, China}

\author{Yanting  Wang}
\affiliation{Institute of Theoretical Physics, Chinese Academy of Sciences, Beijing 100190, China}
\affiliation{School of Physical Sciences, University of Chinese Academy of Sciences, Beijing 100049, China}
\affiliation{Center for Theoretical Interdisciplinary Sciences, Wenzhou Institute, University of Chinese Academy of Sciences, Wenzhou, Zhejiang 325001, China}

\author{Jing Zhou}
\email{zhoujing@dimst.ac.cn}
\affiliation{Dongguan Institute of Materials Science and Technology, Chinese Academy of Sciences, Dongguan 523808, China}
\affiliation{Songshan Lake Materials Laboratory, Dongguan 523808, China}

\author{Haiyang Bai}
\email{hybai@iphy.ac.cn}
\affiliation{Songshan Lake Materials Laboratory, Dongguan 523808, China}
\affiliation{\textit{Institute of Physics, Chinese Academy of Sciences, Beijing 100190, China}}

\author{Yuliang Jin}
\email{yuliangjin@mail.itp.ac.cn}
\affiliation{Institute of Theoretical Physics, Chinese Academy of Sciences, Beijing 100190, China}
\affiliation{School of Physical Sciences, University of Chinese Academy of Sciences, Beijing 100049, China}
\affiliation{Center for Theoretical Interdisciplinary Sciences, Wenzhou Institute, University of Chinese Academy of Sciences, Wenzhou, Zhejiang 325001, China}

\begin{abstract}
Dynamic hysteresis, the rate-dependent lagged response of materials to external fields, underpins applications from energy-efficient transformers to gas storage systems. A fundamental yet unresolved question is how the hysteresis loop area $A$ scales with the field sweep rate $R$. Here, we reveal that a competition between the field sweep and thermal fluctuations governs a universal crossover between two scaling regimes: $A - A_0 \propto R^{1/3}$ for $R < R^*$ and $A - A_0  \propto R^{2/3}$ for $R > R^*$, where $A_0$ is the quasi-static area and the crossover rate $R^* \propto T/T_c$ depends on the temperature $T$ and the material's critical temperature $T_c$. We demonstrate these scaling laws universally across experiments of magnetic materials, simulations of Ising and metal-organic framework models, and analytical solutions of a stochastic Langevin equation. This framework not only resolves the long-standing non-universality of reported scaling exponents but also provides a direct design principle for the application of dynamic  hysteresis.
\end{abstract}

\textit{}
\maketitle

{\bf Introduction.}\label{Intro}
Hysteresis refers to the ubiquitous history-dependent response to an externally applied driving field in systems with an underlying first-order phase transition (FOPT)~\cite{brokate2012hysteresis, krasnosel2012systems,JMagnMagnMater.61.48}.  
In most cases, hysteresis is inherently both {\it dynamical} and {\it thermal}: the delay in the response $\phi$ depends on the rate $R$ of change of the driving field $H$ and the temperature $T$ of the system~\cite{RevModPhys.71.847}. The dependence of the area $A$ of the hysteresis ($\phi$-$H$) loop  on $R$ is known as {\it hysteresis dispersion}.
For {\it dynamical-thermal hysteresis}, this  relationship $A(R)$ is itself a function of $T$.

Hysteresis dispersion is a critical property in a vast range of applications. 
For example, in electric transformers, hysteresis loss (proportional to $A$) must be minimized to reduce energy wasted as heat in magnetic cores~\cite{Science.362.eaao0195,JApplPhys.64.6044}.
In contrast, maximizing $A$ is desirable in other contexts, such as for gas storage in porous materials like metal-organic frameworks (MOFs), where a large, rate-sensitive hysteresis can enhance working capacity~\cite{NatChem.1.695, Science.341.1230444}. 

Despite its importance, a unified understanding of the behavior of $A(R)$ has remained elusive. 
A power-law scaling of the form $A - A_0 \propto R^\alpha$, where $A_0$ is the quasi-static ($R \to 0$) hysteresis area, has been reported in numerous experiments and simulations across diverse systems, including magnetic materials~\cite{PhysRevB.52.14911,PhysRevLett.78.3567,PhysRevLett.70.2336}, electric material~\cite{PhysRevB.55.R11933}, optical devices~\cite{PhysRevLett.65.1873,PhysRevLett.74.2220}, strong correlation electronic systems~\cite{PhysRevLett.121.045701,PhysRevE.108.024101}, cold atomic system~\cite{PhysRevE.94.032141}, Ising models~\cite{PhysRevA.42.7471,RevModPhys.71.847}, and Langevin dynamics~\cite{PhysRevB.42.856,PhysRevE.51.2898}. 
While the data convincingly exhibit scaling over several orders of magnitude, the exponent $\alpha$ is found to be non-universal. A survey of the literature (see Supporting Information (SI) Table~\ref{tab:coefficient}) reveals that $\alpha$ typically falls between 0 and 1, with its distribution showing two prominent peaks centered near $\alpha_1 = 1/3$ and $\alpha_0 = 2/3$ (see SI Fig.~\ref{fig:exponent_distribution}).

The scaling of $A(R)$ is well understood theoretically in the athermal limit ($T\to0$). In this regime, the exponent $\alpha_0 = 2/3$ is successfully obtained from the mean-field (MF) solution of an athermal Langevin equation~\cite{PhysRevLett.65.1873} and from the driven dynamics of the Curie-Weiss model~\cite{PhysRevLett.134.177101}.
What remains poorly understood is how $T$ alters the scaling of  $A(R)$. The framework of \emph{complete universal scaling} (CUS)  proposes collapsing hysteresis loops $\phi(H, R, T, a_2, a_4)$ by rescaling all relevant parameters with appropriate powers of  $R$~\cite{PhysRevLett.95.175701, ChinPhysLett.41.100502}:
$(\phi - \phi_{s}) R^{-\frac1 3}= \Psi\left[ a_2R^{-\frac1 3}, a_4R^{\frac1 3}, (H-H_{s})R^{-\frac2 3}, T R^{-1} \right]$, where $(H_s, \phi_s)$ is the spinodal point of $\Psi (H)$ with $T=R=0$, and $\Psi$ is a scaling function. 
This method requires rescaling parameters $a_2$ and $a_4$ appearing in the MF Landau free-energy.
However, the system's free energy is generally non-adjustable and often unknown in experiments and simulations. Moreover, it does not explicitly give the $T$-dependence of $A(R)$.

Here, we propose a two-piece scaling form that incorporates both \(\alpha_1 = 1/3\) and \(\alpha_0 = 2/3\) exponents:
\begin{equation}
A - A_0 \propto 
\begin{cases} 
R^{1/3}, & R < R^*, \\ 
R^{2/3}, & R > R^*, 
\end{cases}
\label{eq:scaling}
\end{equation}
where,
\begin{equation}
R^* = R_0 \frac{T}{T_c},
\label{eq:R}
\end{equation}
with \(T_{\mathrm{c}}\) denoting the critical temperature at which the FOPT (coexistence) line terminates, and \(R_0\) being a constant. This scaling behavior is observed experimentally in five magnetic materials and supported by simulations of Ising and MOF models. We further derive these scaling laws by combining  the theories of Langevin dynamics  and complete universal scaling. 
Although the exponents are derived by MF theories, they agree fairly well with experimental data. 

Physically, Eqs.~(\ref{eq:scaling}) and~(\ref{eq:R}) result from the intrinsic competition between field sweep and thermal fluctuations. The condition $R/R_0 < T/T_c$ facilitates thermal fluctuations relative to the driving field - the  thermal activation from a metastable state thereby reduces dissipative losses compared to a counterpart athermal system ($T=0$) driven by the same rate $R$.
Conversely, if $R/R_0 > T/T_c$, thermal fluctuations do not occur within the time scale $\sim 1/R$  and the system becomes effectively athermal. 


The universal scaling described by Eqs.~(\ref{eq:scaling}) and~(\ref{eq:R}) leads to a useful principle for material design. Despite extensive research, the nonlinear behavior of $A(R)$ often relies on empirical descriptions~\cite{AIPAdv.15.035115,JApplPhys.7.07A322}, making design a process of trial and error. The above scaling  establishes a straightforward guideline: to minimize (maximize) the hysteresis loss in a system, one should optimize the combination of $R$ and $T$ to reach the thermal scaling regime $R<R^*=R_0(T/T_c)$ (athermal scaling regime, $R>R^*$).



\begin{figure*}[htbp]
\includegraphics[width=0.33\textwidth]{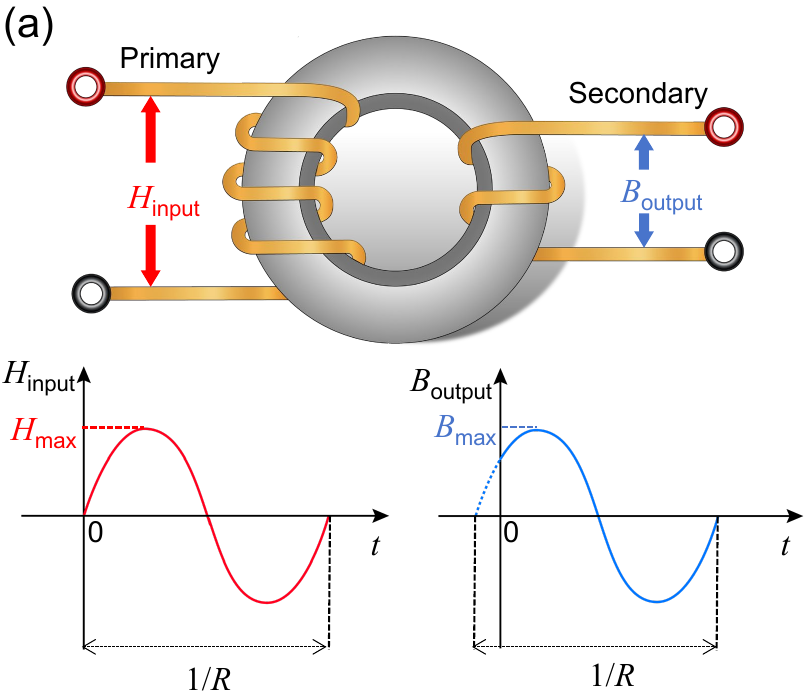}\includegraphics[width=0.67\textwidth]{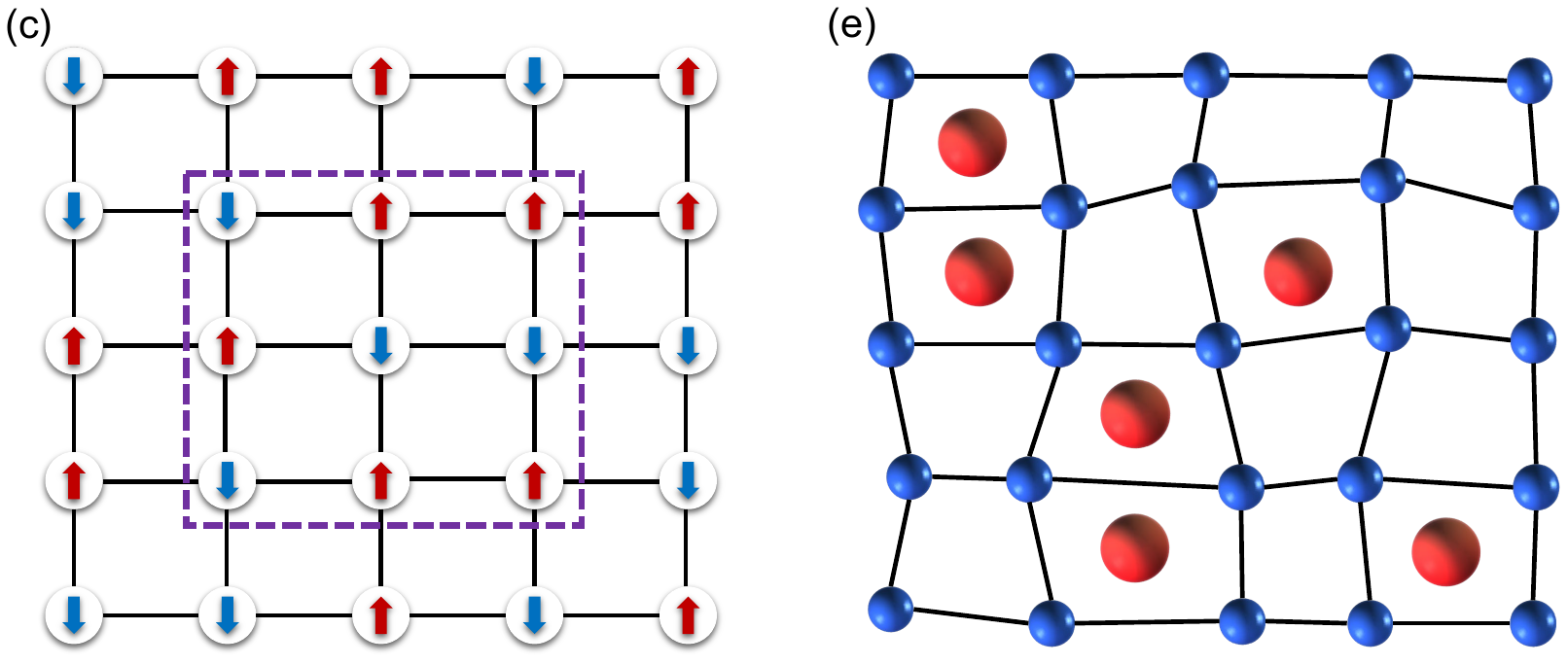}
\includegraphics[width=\textwidth]{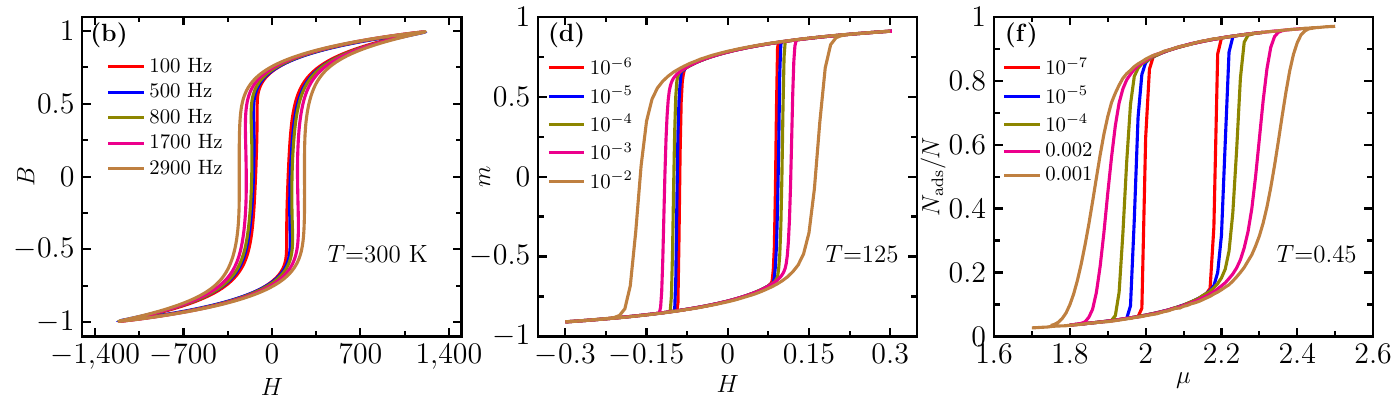}
	\caption{\label{fig:system} 
    {\bf Systems and their hysteresis curves.}
    (a) Illustration of the experimental setup.
    (b) Magnetic induction - magnetic field ($B$-$H$) curves measured in experiments. 
    (c) Illustration of the generalized Ising model with $l=3$ (dashed square).
    (d) Magnetization - magnetic field ($m$-$H$) curves measured in simulations of the generalized Ising model ($l=13$). (e) Illustration of the MOF model. (f) Fraction of guest particles adsorbed in the host MOF, $N_{\rm ads}/N$, as a function of their chemical potential $\mu$.
    In (b,d,f), $R$ values are indicated in the figure legend.
	}
\end{figure*}
{\bf Experiments and simulations.}
We study hysteresis curves in the following systems.

{\it (i) Experiments of magnetic materials.} 
Ferromagnetic materials exhibit dynamic magnetic induction  ($B$) responses  under alternating magnetic field ($H$) excitation, characterized by typical hysteresis behavior~\cite{ApplPhysRev.5.031301,AdvMater.23.821}.
Here, five representative ferromagnetic alloys, namely $\mathrm{Ni_{80}Fe_{15}Mo_{5}}$ permalloy, $\mathrm{Fe_{73.5}Si_{15.5}B_{7}Cu_{1}Nb_{3}}$ nanocrystalline alloy, $\mathrm{Fe_{48}Co_{50}V_{2}}$ alloy, Fe-3wt$\%$Si alloy and MnZn ferrite, are selected and fabricated into toroidal cores. Regulated by the synergistic effects of chemical compositions and microstructures, these materials exhibit  a wide range of  Curie temperatures $T_c$ (see SI Table~\ref{SItab:Temperature}).

Hysteresis loop measurements of the samples are performed using an AC $B$-$H$ analyzer, with the AC frequency $R$ ranging from 50 Hz to 3000 Hz. Prior to the tests, copper wire windings with appropriate turns are wound on both sides of the toroidal core (see Fig.~\ref{fig:system}(a)).
The primary winding is connected to an excitation source to apply alternating magnetic fields with different frequencies, thereby regulating the intensity of the externally applied magnetic field ($H=H_{\mathrm{input}}$). The secondary winding is dedicated to the real-time acquisition of signals reflecting the dynamic variation of the magnetic induction intensity ($B=B_{\mathrm{output}}$) inside the core (see Methods Sec.~A for further experimental details).
To ensure consistency, all samples are magnetized to reach saturation under the same maximum value $H_{\rm max}$ of the applied field intensity.  Typical hysteresis loops are presented in Fig.~\ref{fig:system}(b) (see SI Sec.~\ref{SIsec:experiment} for additional data).
The area $A$ of the hysteresis loop is obtained by integrating the $B$-$H$ curve.  

{\it (ii) Simulations of Ising models.}
We perform Monte Carlo (MC) simulations of standard Ising models with nearest-neighbor (NN) interactions in two and three dimensions (2D and 3D), as well as a generalized Ising model. In the generalized model, the spins ${s_i}$ interact pairwise within an $l \times l$ square (containing $l \times l$ spins) (see Fig.~\ref{fig:system}(c) and Methods Sec.~B). When $l > l^* \approx 10$, this model is effectively MF; see SI Sec.~\ref{SI:Ising}.
At $T<T_{c}$, a thermodynamic FOPT occurs when the sign of $H$ is reversed.
In our simulations, the magnetic field varies linearly with time $t$ (measured in MC steps), $dH/dt = R$, and its direction is reversed ($R \to -R$) when $|H|$ reaches a large preset threshold $H_{\rm max}$. The response, quantified by the magnetization $m = \frac{1}{N} \sum_{i=1}^N s_i$, exhibits standard hysteresis behavior (see Fig.~\ref{fig:system}(d) for the MF Ising model with $l=13$), from which the hysteresis area $A(R)$ is determined (additional data are provided in SI Sec.~\ref{SI:Ising}).

{\it (iii)  Simulations of a MOF model.}
MOFs are soft porous crystals built from metal nodes connected by organic linkers~\cite{NatChem.1.695, Science.341.1230444}. They exhibit promising capabilities of molecular adsorption, which have wide industrial applications, including gas separation, storage, and release~\cite{NatChem.1.695}. We employ the 2D coarse-grained square lattice model
incorporating elastic heterogeneity~\cite{PNAS.120.2302561120} to simulate the  adsorption-desorption hysteresis in MOFs using the MC method (see Fig.~\ref{fig:system}(e) and Methods Sec.~C for details). The fraction of adsorbed molecules $N_{\rm ads}/N$ shows  hysteresis when the chemical potential $\mu$ of the guest particle changes alternately (see Fig.~\ref{fig:system}(f)). The hysteresis area $A(R)$ is estimated from the $N_{\rm ads}/N$-$\mu$ curve for the given linear rate $R$ (see SI Sec.~\ref{SI:MOF} for additional results).

\begin{figure}[htbp]
	\includegraphics[width=\columnwidth]{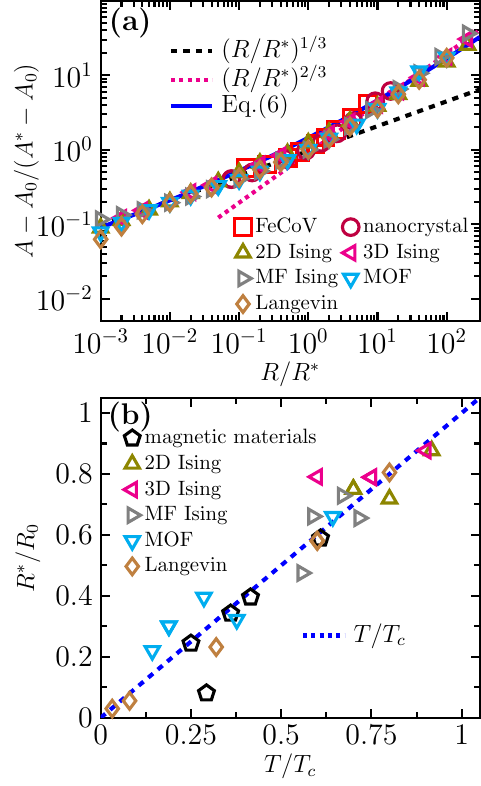}
	\caption{\label{fig:scaling} 
    {\bf Universal scaling of hysteresis  dispersion in experiments and simulations.}
    (a) $(A-A_0)/(A^*-A_0)$ as a function of $R/R^*$, where $A^* \equiv A(R^*)$. We include results from the Langevin equation, MOF models, Ising models and two experimental magnetic materials  (nanocrystalline and FeCoV alloy). 
    (b) $R^*/R_0$ as a function of $T/T_c$ (see SI Table~\ref{SItab:Temperature} for the list of $T$, $T_c$ and other model parameters).
	}
\end{figure}

The above  systems universally obey the scaling Eqs.~(\ref{eq:scaling}) and (\ref{eq:R}), as shown in Fig.~\ref{fig:scaling}. 
In simulations, the quasi-static area $A_0$ is estimated by the plateau value of $A(R)$ in the small-$R$ limit. In experiments, the quasi-static limit is inaccessible, and $A_0$ is extrapolated from the finite-$R$ data (see SI Sec.~\ref{SI:A_0}).  
Once $A_0$ determined, the $A(R)$ data exhibit  $A-A_0 \propto R^{1/3}$ for $R<R^*$, and exhibit  $A-A_0 \propto R^{2/3}$ for $R>R^*$, where $R^*$ is the crossover rate separating  the two behaviors. The data obtained from different systems all collapse, if they are rescaled by the crossover values $(A^*, R^*)$ (see Fig.~\ref{fig:scaling}(a)). Furthermore, $R^*$ obeys Eq.~(\ref{eq:R}), as shown by  Fig.~\ref{fig:scaling}(b). The constant $R_0$  is universal within each category: (i) $R_0 \approx 1550$ Hz for experimental magnetic materials, (ii)  $R_0 \approx 0.056$ for Ising models, and (iii) $R_0 \approx 0.001$ for the MOF model. In fact, $1/R_0$ can be simply understood as the unit of time.

{\bf Langevin dynamics.}\label{sec:Langevin}
A standard theoretical description of hysteresis is based on the Langevin equation  of the order parameter $\phi$~\cite{RevModPhys.49.435,goldenfeld2018lectures}:
\begin{align}
	\frac{d \phi}{d t} = -\lambda \frac{d \mathcal{F}(\phi)}{d \phi} + \xi,
	\label{eq:LE}
\end{align}
in which $\lambda$ is a kinetic constant and $\xi$ is a Gaussian white noise, i.e., $\langle \xi(t)\rangle=0$ and $\langle \xi(t)\xi(t')\rangle = 2 \lambda T \delta(t-t')$ (the Boltzmann constant $k_{\rm B} = 1$). Here we consider the MF Landau free energy with a time-dependent external field $H = Rt$, 
\begin{align}
\mathcal{F}(\phi) = \frac{1}{2}a_2 \phi^2 + \frac{1}{4}a_4 \phi^4 - H\phi,
\label{eq:FreeEnergy}
\end{align}
where $a_2$ and $a_4$ are constants, and in general one sets $a_4>0$. It is well known that a continuous transition occurs at $a_2 = 0$, while  $a_2 < 0$ corresponds to a field-driven FOPT. Substituting Eq.~(\ref{eq:FreeEnergy}) into Eq.~(\ref{eq:LE}), we obtain,
\begin{align}
	\frac{d \phi}{d t} = -\lambda (a_2\phi + a_4\phi^3-H) + \xi.
	\label{eq:LE2}
\end{align}

\begin{figure}[htbp]
	\includegraphics[width=\columnwidth]{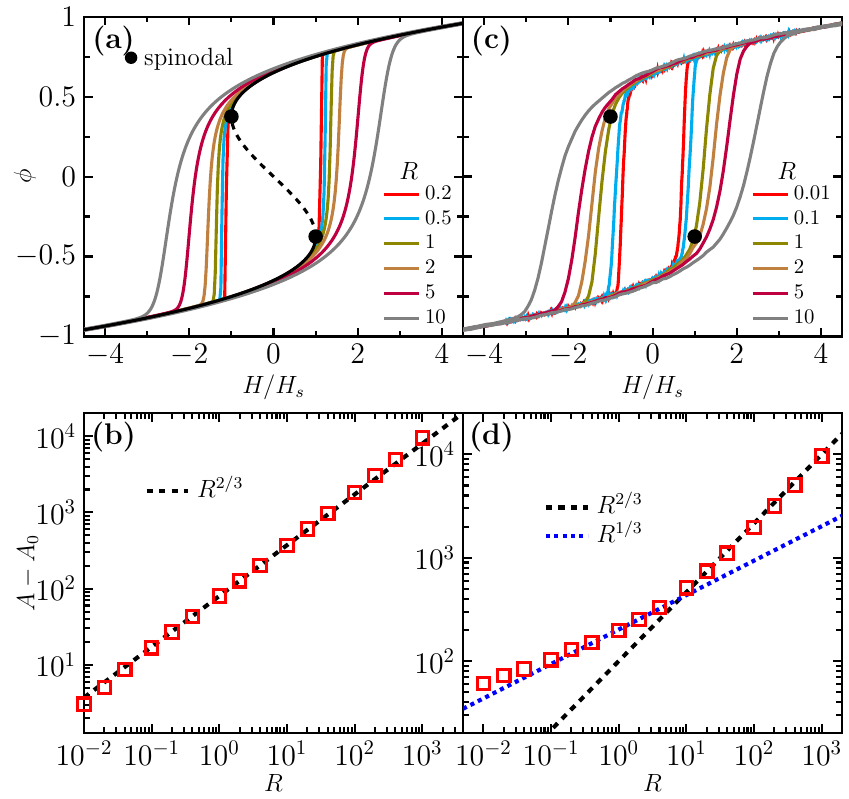}
	\caption{\label{fig:LEsigma}
    {\bf Langevin dynamics.}
    (a) Dynamical hysteresis by solving Eq.~(\ref{eq:LE2}) with $T=0$. The solid (dashed) black curve is the quasi-static equation $H=a_2\phi+a_4\phi^3$, and the two solid circles denote the spinodal points. (b) Hysteresis area $A-A_0$ versus $R$ with $T = 0$.
    (c) Dynamical-thermal hysteresis for Eq.~(\ref{eq:LE2}) with $T=2$ and its corresponding scaling (d).
    We set $a_2=-400$, $a_4=948$ and $\lambda=0.0005$. 
	}
\end{figure}

With the noise term ($T=0$) ignored in Eq.~(\ref{eq:LE2}), previous studies have established, both analytically~\cite{PhysRevLett.65.1873,PhysRevE.108.024101} and numerically~\cite{PhysRevLett.65.1873,PhysRevE.50.224}, a scaling law, $A - A_0 \propto R^{\frac2 3}$. This can be seen from the following analysis. First, the stationary ($R \to 0$) solution $H = a_2\phi + a_4\phi^3$ of Eq.~(\ref{eq:LE2}) is obtained  by setting $\frac{d \phi}{d t} = 0$ (the black line  in Fig.~\ref{fig:LEsigma}(a)). This solution is unstable between the two spinodal points, $(H_s,\phi_s)=(\frac{2a_2}{3}\sqrt{\frac{-a_2}{3a_4}}, \sqrt{\frac{-a_2}{3a_4}})$ and $(-H_s,-\phi_s)$, determined by  $\frac{d H}{d \phi} = 0$. With an increasing $R$,  the hysteresis loop becomes lager, as shown by the numerical solutions in Fig.~\ref{fig:LEsigma}(a). In this case,  the change in  $A$ scales with $R$, following $A - A_0 \propto R^{\frac2 3}$ (see Fig.~\ref{fig:LEsigma}(b)), where $A_0$ is the area of the stationary loop.

If $T>0$, thermal fluctuations can activate the system and drive it to overcome the free energy barrier. Numerical solutions of Eq.~(\ref{eq:LE2}) show that the jump of $\phi$ can occur before the spinodal point (see Fig.~\ref{fig:LEsigma}(c)).  The data of the hysteresis area reveals a new scaling, $A - A_0 \sim R^{1/3}$, for $R<R^*$ (see Fig.~\ref{fig:LEsigma}(d)), consistent with the experimental and simulation observations in Fig.~\ref{fig:scaling}(a).
Varying $T$, we find that $R^* \sim T$, consistent with Eq.~(\ref{eq:R}), as shown in Fig.~\ref{fig:scaling}(b), where $R_0/T_c$ is a fitting parameter.

{\bf Scaling theory.} To explain the above results, we develop a scaling theory.  It can be shown that the hysteresis area is described by a scaling form,
\begin{align}
	A-A_0
    \propto  R^{\frac{2}{3}}\left[ 1+c_0\left(\frac{T}{R} \right)^{\frac{1}{3}} \right],
	\label{eq:CorrectScaling21}
\end{align}
where $c_0$ is a  $T$-independent constant. This form fits our experimental and simulated data well (see Fig.~\ref{fig:scaling}(a)).
When $R\rightarrow\infty$, the second term tends to zero and Eq.~(\ref{eq:CorrectScaling21}) reduces to the athermal result, $A - A_0 \propto R^{2/3}$, whereas when $R\rightarrow0$, the second term dominates and Eq.~(\ref{eq:CorrectScaling21}) becomes $A-A_0 \propto R^{1/3}$. The separation of the two scalings occurs at $R^* \sim T$. 
To make the variables dimensionless, we rescale $T$ by  the natural reference temperature $T_c$, and $R$ by a characteristic  rate $R_0$, which is a fitting parameter.  Thus Eq.~(\ref{eq:CorrectScaling21}) is consistent with Eqs.~(\ref{eq:scaling}) and~(\ref{eq:R}). 

\begin{figure}[htbp]
\includegraphics[width=\columnwidth]{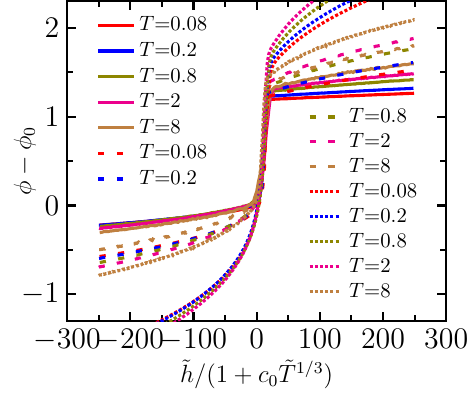}    
	\caption{{\bf Scaling collapse according to Eq.~(\ref{eq:CUS21}).} Data are obtained by numerically solving the Langevin equation Eq.~(\ref{eq:LE2}), with $a_2=-400$, $a_4=948$ and $\lambda=0.0005$. The fitting parameter $c_0=4$. Solid lines, dashed lines and dotted lines correspond to $R=0.01$, $R=1$ and $R=100$, respectively. 
    }
\label{fig:scaling_collapse}
\end{figure}

Eq.~(\ref{eq:CorrectScaling21}) suggests a scaling function for the order parameter (see SI Sec.~\ref{SI:derivation} for the derivation), 
\begin{align}\label{eq:CUS21} 
    \phi - \phi_0 = 
    \Phi(\tilde{h}_{\rm eff}) = \Phi \left[\frac{\tilde{h}}
    {1+c_0\tilde{T}^{\frac1 3}} \right],
\end{align}
where $\tilde{h}\equiv(H-H_0)R^{-\frac2 3}$, $\tilde{T}\equiv T R^{-1}$ and $\tilde{h}_{\rm eff} = \frac{\tilde{h}}{1+c_0\tilde{T}^{\frac1 3}}$.
We can show that Eq.~(\ref{eq:CUS21}) is equivalent to Eq.~(\ref{eq:CorrectScaling21}).
Indeed, because the height  (along the $\phi$-axis) of the hysteresis loop is independent of $R$ (see Fig.~\ref{fig:LEsigma}(a), (c)), we only need to consider the $R$-dependence of $H$.
Then, $A-A_0 \propto H-H_0 \propto R^{\frac 2 3}\left[\tilde{h}_{\rm eff}(1+c_0 \tilde{T}^{\frac 1 3}) \right]$. This is equivalent to Eq.~(\ref{eq:CorrectScaling21}), considering that   $\tilde{h}_{\rm eff}$ is a constant for a given $\phi$ according to Eq.~(\ref{eq:CUS21}). Eq.~(\ref{eq:CUS21}) is validated by the numerical solutions of the Langevin equation in Fig.~\ref{fig:scaling_collapse}, through the data collapse in the transition regime where  $A$ is determined. 

It is interesting to compare Eq.~(\ref{eq:CUS21}) with the CUS proposed in Refs.~\cite{PhysRevLett.95.175701,ChinPhysLett.41.100502},
$\tilde{\phi} = \Psi (\tilde{a}_2, \tilde{a}_4, \tilde{h}, \tilde{T})$,
where $\tilde{\phi}\equiv(\phi-\phi_0)R^{-\frac1 3}$,  $\tilde{a}_2  \equiv a_2R^{-\frac1 3}$ and $\tilde{a}_4  \equiv a_4R^{\frac1 3}$. While Eq.~(\ref{eq:CUS21}) and CUS rescale $T$ and $H$ in the same way, 
there are three essential differences. (i) The CUS requires to rescale all relevant parameters to collapse the hysteresis curves, including the potential energy parameters $a_2$ and $a_4$ in Eq.~(\ref{eq:FreeEnergy}), which are  fixed in our experiments and simulations. (ii) Eq.~(\ref{eq:CUS21}) shows that $\tilde{h}$ and $\tilde{T}$ are not independent as in the CUS; they should be combined into a single variable, $\tilde{h}_{\rm eff}$ - this results in  the crossover between  $1/3$ and $2/3$ scalings. (iii) While the CUS collapses entire hysteresis loops, Eq.~(\ref{eq:CUS21}) only collapses them in the transition regime, $0 \leq \phi-\phi_0 \leq 2(\phi_I - \phi_0)$, where $(H_I, \phi_I)$ is the inflection point at the maximum $d\phi/dH$. Indeed, a similar scaling is observed for $H_I$: $H_I - H_0 \propto R^{1/3}$ for $R<R^*$, and $H_I - H_0 \propto R^{2/3}$ for $R>R^*$ (see SI Sec.~\ref{SI:derivation}). Because $A - A_0 \propto H_I - H_0$ (the height of $A$ is $R$-independent), this scaling is fully consistent with Eq.~(\ref{eq:scaling}).

The above scaling functions, Eqs.~(\ref{eq:CorrectScaling21}) and~(\ref{eq:CUS21}), reveal the origin of the universal scaling of dynamical-thermal  hysteresis,  Eqs.~(\ref{eq:scaling}) and~(\ref{eq:R}). First, in the athermal limit $T \to 0$, the exponent $\alpha_0 = 2/3$ is solely determined by the $R$-dependent scaling of the field, $\tilde{h} = (H-H_0)R^{-\frac2 3}$, which is a dynamical effect.
Second, in the high temperature regime, the scaling invariable 
 becomes, $\tilde{h}/\tilde{T}^{1/3} = (H-H_0)(R T)^{-1/3}$, giving the $1/3$ scaling. Third, the crossover frequency $R^*$ is $T$-dependent because of the invariance $\tilde{T}\equiv T R^{-1}$. If  $R > R^*$, the dynamics are effectively athermal - it means that the external field is varied  so fast such that activations do not have sufficient time to occur. Our results thus unveil coupled dynamical and thermal effects on  hysteresis.

{\bf Discussion.} 
Our study reveals a fundamental competition between the rate of the applied sweeping field  and thermal fluctuations in determining hysteresis behavior.
These results lead to a practical design principle: one can tune the hysteresis loss through  effectively switch on or off the  thermal fluctuations, by setting $R/R_0 < T/T_c$ or $R/R_0 > T/T_c$ respectively.

The theoretical analyses presented above are based on the MF Langevin equation. The MF exponents ($\alpha_0 = 2/3$ and $\alpha_1 = 1/3$) show good agreement with data from non-MF systems, including the 2D and 3D Ising models and experimental magnetic materials. This agreement suggests that spatial fluctuations do not play a significant role in determining the scaling behavior of hysteresis. In contrast to typical critical phenomena, finite-size effects on the hysteresis loop are negligible (see SI Sec.~\ref{SI:Ising}), implying that spatial correlations are irrelevant. Nevertheless, we cannot exclude the possibility of non-MF corrections to the scaling exponents~\cite{PhysRevLett.95.175701}, as such corrections may be too small to detect given the resolution of our simulations and experiments.

The established universal scaling serves as a solid starting point for more complete descriptions of  hysteresis. For instance, eddy currents - induced in conductive magnetic materials according to Faraday's law of induction - lead to energy losses that scale with the rate as $A(R) \sim R^2$ in the high-frequency regime~\cite{cullity2011introduction}. Such effects, however, are highly system-specific and lie beyond the scope of the present analysis. A promising direction for future research involves investigating how structural disorder influences hysteresis dispersion in amorphous  systems, such as metallic glasses. In these materials, the absence of long-range atomic order can significantly alter domain-wall dynamics and dissipation mechanisms, leading to  more complex behavior such as scale-free avalanches~\cite{nature.410.242}.

{\bf Acknowledgments.}
This work was supported by the National Natural Science Foundation of China (Grant Nos. 52192601, 52192602, 12447101) and  Wenzhou Institute (Grant No. WIUCASICTP2022).
YJ acknowledges funding from  the National Key R\&D Program of China (Grant No. 2025YFF0512000) and Project supported by the Space Application System of China Manned Space Program. 
YS acknowledges funding from National Natural Science Foundation of China (Grant No. 12447118).
The authors acknowledge the use of the High Performance Cluster at Institute of Theoretical Physics, Chinese Academy of Sciences,  and the computer clusters at the Hefei advanced computing center.

\vspace{1cm}
{\Large \bf Methods}
\vspace{0.5cm}

{\bf A. Experimental setup and samples.}
    Commercial $\mathrm{Ni_{80}  Fe_{15}Mo_5}$ permalloy ribbon (100 \si{\um} in thickness, 15 mm in width) is ultrasonically cleaned in anhydrous ethanol for 15 minutes to eliminate surface oils and contaminants. A specialized automatic winding machine is then employed to fabricate toroidal cores with a standard geometry (outer diameter: 30 mm, inner diameter: 20 mm). Using identical cleaning and winding protocols, commercial $\mathrm{Fe_{48}Co_{50}V}$ alloy ribbon (100 \si{\um} in thickness, 15 mm in width) are wound into toroidal cores (outer diameter: 30 mm, inner diameter: 20 mm), whereas Fe-3wt\%Si alloy ribbon (0.4 mm in thickness, 20 mm in width) forms toroidal cores with an outer diameter of 38 mm and an inner diameter of 22 mm. The commercial Mn-Zn ferrite is fabricated via the conventional sintering process, with the resulting sample having dimensional parameters of 36.5 mm $\times$ 15.5 mm $\times$ 23 mm (outer diameter $\times$ inner diameter $\times$ height). For the $\mathrm{Fe_{73.5}Si_{15.5}B_{7}Cu_{1}Nb_{3}}$ nanocrystalline alloy, a thinner and narrower ribbon (15 \si{\um} in thickness, 10 mm in width) is wound into toroidal cores matching the aforementioned geometry, followed by transverse magnetic field crystallization annealing to optimize soft magnetic properties. Specifically, the nanocrystalline core is isothermally annealed at 833 K for 120 minutes under a constant transverse magnetic field of 1000 Gs throughout the annealing process.
    
    The hysteresis loops of all samples are measured using a SY-8258 AC $B$-$H$ analyzer (Iwasaki Tsushinki Co., Ltd., Japan). Prior to the  measurements, the instrument is calibrated with a standard sample to ensure the accuracy of the test results. Subsequently, primary and secondary windings with an appropriate number of turns are wound on each toroidal core, which are then connected to the corresponding interfaces of the analyzer. After inputting the core dimensional parameters and preset test conditions (including the AC frequency and maximum applied field intensity), the hysteresis loops of the materials under diverse frequencies and applied magnetic fields are automatically recorded by the analyzer. All measurements are carried out at the room temperature (about 300 K).

{\bf B. Simulation method of Ising models.}
In the standard Ising model, spins are placed on 2D and 3D square lattices, and interact with their nearest neighbors. 
In addition, we consider  a generalized Ising model on a $L\times L=140\times 140$ square lattice in 2D, whose interaction range is within a $l\times l$ square, as shown in Fig.~\ref{fig:system}(c). The Hamiltonian of this model is given by
\begin{align}
	\mathcal{H}_{\mathrm{Ising}} = -J\sum_{i\forall,j\in l^2}s_is_j - H(t)\sum_i s_i,
	\label{eq:IsingHamil}
\end{align}   
where $l^2 = 3\times 3, 5\times 5, 7\times 7 \ldots$ denotes the $l\times l$ square centered around $i$; $l=3$ is equivalent to the next nearest-neighbor (NNN) Ising model. Here $J\equiv J_0/k_{\rm B}T$ is the coupling strength of interactions and $H(t)\equiv H_0(t)/k_{\rm B}T$ (they are both dimensionless), and $J_0$ and the Boltzmann constant $k_{\rm B}$ are both set to unity hereafter. Thus $J$ 
is the control parameter of the ratio between potential and thermal energies, 
and the other control parameter is the driving rate $R$. In MC simulations, the  magnetic field varies linearly with time $t$ (MC steps), $dH/dt =  R$, with its direction reversed ($R \to -R$) when $|H|$ reaches a large preset threshold $\pm H_{\mathrm{max}}$, to ensure that the hysteresis loop is saturated (see Fig.~\ref{fig:system} (d)).
Before the hysteresis measurement,  
an equilibrium state is prepared by $10^4$ steps  under $H=-H_{\rm max}$ and $T<T_c$, starting from an all-spin-down initial configuration. 
The order parameter, the magnetization $m=\frac{1}{N}\sum_i^{N}s_i$, and the corresponding susceptibility $\chi$, are averaged over 320 independent samples. Periodic boundary conditions are applied in both  directions. 

{\bf C. Simulation method of the MOF model.}
The MOF model~\cite{PNAS.120.2302561120} is a coarse-grained 2D lattice with $N=L\times L$ host particles, where guest particles can be adsorbed/desorbed by the host matrix formed by the host particles, see Fig.~\ref{fig:system}(e). Each plaquette (unit cell) can accommodate only one guest particle, and the plaquette expands/contracts isotropically by adsorbing/desorbing the guest particles. There are  $N$ plaquettes, which can adsorb up to $N$ guest particles.
Periodic boundary conditions are applied.  The Hamiltonian comprises the lattice site positions $\bm{r}_i$ ($i=1, 2, 3, ..., N$) and the variables on the plaquettes $\sigma_j$ ($j=1, 2, 3, ..., N$) taking 1 (presence of a guest particle) or 0 (absence), 
\begin{align}
	\mathcal{H}_{\mathrm{MOF}} =  \sum_{j=1}^N \biggl[V_1(\{\bm{r}_{i\in j}\})+\sigma_j[V_2(\{\bm{r}_{i\in j}\})-\mu] \biggr],
	\label{eq:MOFHamil}
\end{align}   
where $i\in j$ denotes all particles at the vertices of the $j$-th plaquette and $\mu$ is the chemical potential of the adsorption of guest
 particles.
Here, 
\begin{equation}
V_1(\{\bm{r}_{i\in j}\})=\frac1 4 \sum_{\mathrm{NN}}(1-r_{ij})^2+\frac1 2 \sum_{\mathrm{NNN}}(\sqrt{2}-r_{ij})^2
\end{equation}
represents the potential energy of a plaquette, where $r_{ij}$ is the distance between two host particles.
The second term,
\begin{equation}
\begin{split}
&V_2(\{\bm{r}_{i\in j}\}) = \\
&k\left[ \frac1 4 \sum_{\mathrm{NN}}(1+a-r_{ij})^2+\frac1 2 \sum_{\mathrm{NNN}}(\sqrt{2}(1+a)-r_{ij})^2 \right],
\end{split}
\end{equation}
is the effective interaction between the host and guest particles, where $k$ is the relative energy scale of the host-guest interaction, and $a$ is a deformation coefficient of the plaquette. Here the elastic constant and natural length of the square plaquette are both set to unity, and thus all quantities are nondimensionalized. 
When $ka>0$, the adsorption gives rise to lattice expansion; whereas when $ka<0$, it induces contraction. We only focus on the case of $ka>0$. We set $L=24$, and have examined that the finite size effects are negligible.

Since we consider isotropic swelling in the adsorption process, the osmotic ensemble~\cite{JAmChemSoc.130.14294} is appropriate.
The control thermodynamic parameters are the hydrostatic pressure $P$, temperature $T$ and  chemical potential of guest particles $\mu$. Here we set $P=0$ without losing generality, and fix $T$ below $T_c$ to study the adsorption-desorption hysteresis  caused by  changing $\mu$. We use two sets of parameters, $k=5$, $a=0.6$ and $k=3$, $a=2/3$, with critical temperatures  $T_c = 0.7$ and 0.53, respectively.
The corresponding phase diagrams  are presented in Ref.~\cite{PNAS.120.2302561120}.

MC simulations are performed to investigate the hysteresis behavior of the adsorption-desorption transition. Firstly, in the initial configuration,   all plaquettes are unoccupied ($\left\{ \sigma_i=0 \right\}$). Next we equilibrate the system by $t=10^4$ MC steps under the chemical potential $\mu=\mu_{\mathrm{min}}$. 
Then $\mu = R t$ varies linearly (similar to the Ising MC simulations in Methods Sec. B).
 Each MC step consists of three parts: one Metropolis sweep for the occupied plaquettes  $\left\{ \sigma_i \right\}$, $L$ iterations of sweeps for the lattice positions $\bm{r}_i$, and $L$ iterations for the affine volume $V$. The average swelling ratio is $b\equiv \sqrt{V/V_0}$, where $V_0=N$. The MC step sizes are limited such that $|\bm{r}_i|<0.1$ and $|b|<0.01$, and the bond overlaps are prohibited to preserve the square lattice configuration. The order parameter is the fraction of guest particles, $N_{\rm ads}/N$, which is averaged over 320 independent samples.

\bibliography{mybib}

\clearpage
\onecolumngrid
\section{Summary of the coefficient $\alpha$ reported in the literature.}
\begin{figure*}[htbp]
\includegraphics[width=0.5\columnwidth]{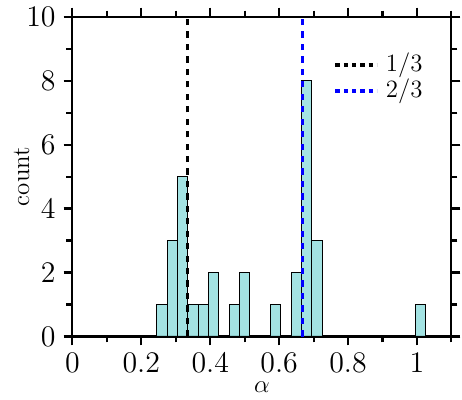}    
	\caption{\bf Histogram of the exponent $\alpha$ reported in the literature (see Table~\ref{tab:coefficient} for the list). 
    }
\label{fig:exponent_distribution}
\end{figure*}

The exponent $\alpha$ in the scaling $A-A_0 \propto R^\alpha$ has been extensively studied. Table~\ref{tab:coefficient} provides a list of $\alpha$ reported in the literature. In Fig.~\ref{fig:exponent_distribution}, the histogram of these reported values is plotted.

\begin{table*}[htbp]
\centering
\caption{{Coefficient $\alpha$ reported in the literature.} 
\label{tab:coefficient}
}
\begin{tabular}{|c|c|c|c|}
\hline
\textbf{Category} & \textbf{System} & \textbf{Method} & \textbf{Scaling exponent} \\
\hline
\multirow{3}{*}{magnetic materials} & Co/Cu film~\cite{PhysRevB.52.14911} & \multirow{19}{*}{experimental} & 2/3\\
\cline{2-2}\cline{4-4}
 & Fe/W film~\cite{PhysRevLett.78.3567} & & 0.02\\
 \cline{2-2}\cline{4-4}
& Fe/Cu film~\cite{PhysRevLett.70.2336} & & 0.31\\
\cline{1-2}\cline{4-4}

\multirow{4}{*}{electric materials} 
 & PZT thin film~\cite{JApplPhys.86.5198,MaterLett.52.213} & & 1/3\\
 \cline{2-2}\cline{4-4}
& BPA bulk~\cite{PhysRevB.55.R11933} & & 0.4\\
\cline{2-2}\cline{4-4}
& SBT thin film~\cite{ApplPhysLett.83.1406} & & 2/3\\
 \cline{2-2}\cline{4-4}
& PLT~\cite{MaterSciEngB.99.179} & & 1/3\\
\cline{1-2}\cline{4-4}

\multirow{3}{*}{Mott transition} 
 & VO$_2$~\cite{SolidStateCommun.97.847} & & 0.42\\
\cline{2-2}\cline{4-4}
 & V$_2$O$_3$~\cite{PhysRevLett.121.045701} & & 0.62(heat) 0.64(cool)\\
 \cline{2-2}\cline{4-4}
& NdNiO$_3$ thin film ~\cite{PhysRevE.108.024101} & & 0.36(heat) 0.31(cool)\\
\cline{1-2}\cline{4-4}

\multirow{2}{*}{Martensitic transition} 
& Co~\cite{SolidStateCommun.114.231} &  & 0.39(heat) 0.49(cool)\\
 \cline{2-2}\cline{4-4}
 & FeMn alloy~\cite{SolidStateCommun.97.847} & & 0.699\\
\cline{1-2}\cline{4-4}

\multirow{2}{*}{structure transition} 
& RbNO$_3$ ~\cite{SolidStateCommun.97.847} & & 0.28\\
 \cline{2-2}\cline{4-4}
 & liquid crystal~\cite{PhysRevE.69.031705} & & 0.692\\
\cline{1-2}\cline{4-4}

\multirow{1}{*}{cold atomic system} 
& Rb atom~\cite{PhysRevE.94.032141} & & 0.64\\
\cline{1-2}\cline{4-4}

\multirow{6}{*}{optical device} 
 & bistable semiconductor laser diode~\cite{PhysRevLett.65.1873} & & 2/3\\
 \cline{2-2}\cline{4-4}
& bistable semiconductor laser ~\cite{PhysRevLett.74.2220} & & 2/3\\
\cline{2-2}\cline{4-4}
 & optical cavity~\cite{PhysRevLett.124.153603} & & 0.25\\
 \cline{2-2}\cline{4-4}
& bistable semiconductor laser~\cite{PhysRevLett.118.247402} & & 1.0\\
\cline{2-4}
& photonic resonators~\cite{PhysRevA.93.033824} & \multirow{12}{*}{numerical} & 2/3\\
 \cline{2-2}\cline{4-4}
& bistable semiconductor laser~\cite{PhysRevLett.74.2220} & & 0.6\\
\cline{1-2}\cline{4-4}

\multirow{6}{*}{Ising model} 
 & 2D Ising~\cite{PhysRevA.42.7471,PhysRevB.52.6550} &  & 0.36\\
 \cline{2-2}\cline{4-4}
& 2D Ising~\cite{RevModPhys.71.847} & & 0.3\\
\cline{2-2}\cline{4-4}
 & 3D Ising~\cite{PhysRevB.52.6550} & & 0.45\\
 \cline{2-2}\cline{4-4}
& 3D Four-spin Ising (SC)~\cite{JPhys.10.275} & & 0.47\\
\cline{2-2}\cline{4-4}
& 3D Four-spin Ising (FCC)~\cite{JPhys.10.275} & & 0.7\\
\cline{2-2}\cline{4-4}
& mean-field Ising~\cite{PhysRevE.50.224} & & 2/3\\
\cline{1-2}\cline{4-4}

\multirow{2}{*}{Heisenberg model} 
 & 3D mean-field Heisenberg~\cite{PhysRevB.65.014416} &  & 0.25\\
 \cline{2-2}\cline{4-4}
& 2D mean-field Heisenberg~\cite{PhysRevB.65.014416} & & 0.25\\
\cline{1-2}\cline{4-4}

\multirow{6}{*}{Langevin equation} 
& $(\Phi^2)^2$ model~\cite{PhysRevB.42.856} &  & 1/3\\
\cline{2-2}\cline{4-4}
& $(\Phi^2)^2$ model ~\cite{JPhys.6.7785} & & 0.5\\
\cline{2-2}\cline{4-4}
 & $(\Phi^2)^3$ model~\cite{JPhys.6.7785} & & 0.7\\
 \cline{2-2}\cline{4-4}
& $(\Phi^2)^3$ model~\cite{PhysRevE.51.2898} & & 2/3\\
\cline{2-4}
 
& mean field~\cite{PhysRevLett.65.1873,PhysRevE.108.024101} & \multirow{2}{*}{analytical} & 2/3\\
\cline{2-2}\cline{4-4}
& $(\Phi^2)^2$ model ~\cite{JPhysA.25.4967,PhysRevLett.70.3279} & & 0.5\\
\hline

\end{tabular}
\end{table*}

\clearpage
\section{List of $T$ and $T_c$ for studied systems}
Table~\ref{SItab:Temperature} provides a summary of  $T$ and $T_c$ values, for all experimental and numerical systems investigated in Fig.~\ref{fig:scaling}.

\begin{table*}[htbp]
\centering
\caption{\bf{List of $T$ and $T_c$ for studied systems.} 
\label{SItab:Temperature}
}
\begin{tabular}{|c|c|c|}
\hline
\textbf{System} & \textbf{$T$} & \textbf{$T_c$} \\
\hline
MnZn ferrite
& 300 K & 493 K\\
\hline
permalloy & 300 K  &  683 K\\
\hline
nanocrystalline alloy & 300 K & 823 K\\
\hline
Fe-3wt\%Si alloy
& 300 K & 1023 K\\
\hline
FeCoV alloy & 300 K & 1253 K\\
\hline
\multirow{3}{*}{2D Ising model} 
& 1.588& 2.269 \\
\cline{2-3}
& 1.815 & 2.269 \\
\cline{2-3}
& 2.08& 2.269 \\
\hline
\multirow{3}{*}{3D Ising model} 
& 2.703& 4.51 \\
\cline{2-3}
& 3.378 & 4.51 \\
\cline{2-3}
& 4.065& 4.51 \\
\hline
\multirow{4}{*}{MF Ising model} 
& 125& 159.5 ($l=13$)\\
\cline{2-3}
& 250 & 349 ($l=19$)\\
\cline{2-3}
& 333.3&597 ($l=25$)\\
\cline{2-3}
& 500& 597 ($l=25$)\\
\hline
\multirow{5}{*}{MOF model} 
 & 0.1 & 0.53 ($k=3, a=2/3$)\\
 \cline{2-3}
& 0.2 & 0.53 ($k=3, a=2/3$)\\
\cline{2-3}
& 0.1 & 0.7 ($k=5, a=0.6$)\\
\cline{2-3}
& 0.2 & 0.7 ($k=5, a=0.6$)\\
\cline{2-3}
& 0.45 &0.7 ($k=5, a=0.6$)\\
\hline

\end{tabular}
\end{table*}

\clearpage

\section{Additional experimental results}
\label{SIsec:experiment}

\begin{figure}[htbp]
	\includegraphics[width=\columnwidth]{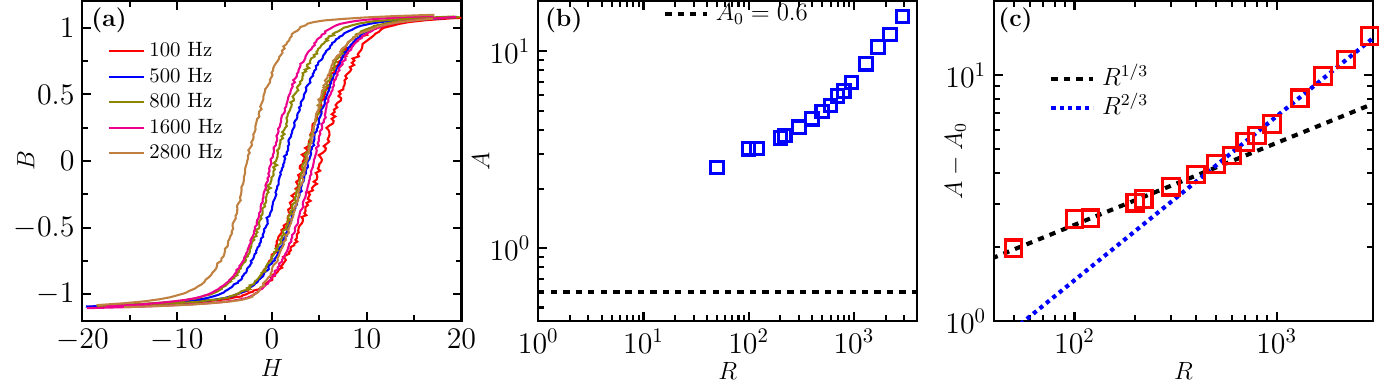}
	\caption{\label{SIfig:nanocrystal} 
    {\bf Experimental results for the nanocrystalline alloy.}
    (a) Magnetic induction - magnetic field ($B$-$H$) curves measured in experiments of nanocrystalline alloy, for different $R$ (indicated in the figure legend). (b) $A$ as a function of $R$  with $A_0=0.6$. (c) Corresponding excess hysteresis area $A-A_0$ versus $R$.  
	}
\end{figure}

\begin{figure}[htbp]
	\includegraphics[width=\columnwidth]{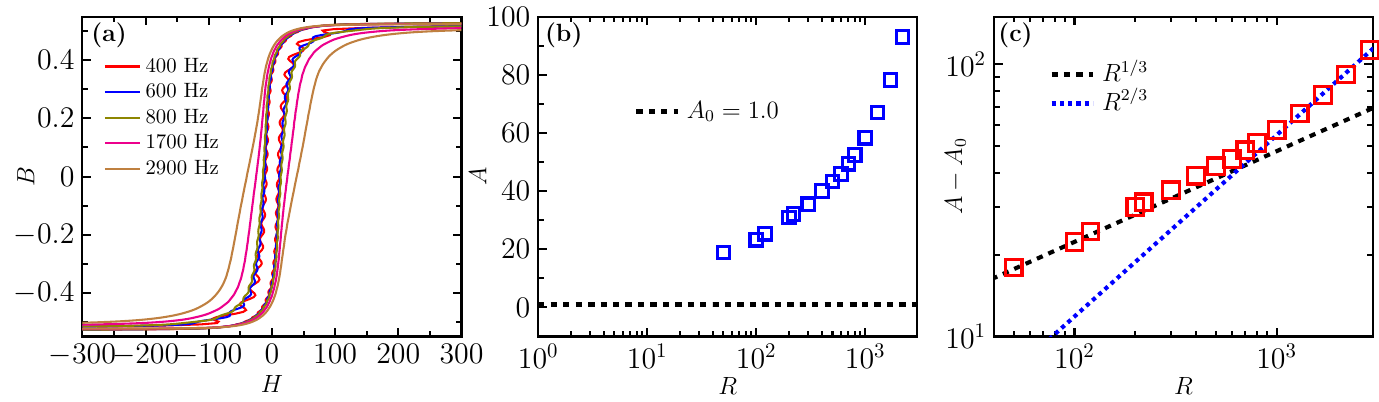}
	\caption{\label{SIfig:permalloy} 
        {\bf Experimental results for the permalloy.}
    (a) $B$-$H$ curves measured in experiments of permalloy. (b) $A$ as a function of $R$ with $A_0=1.0$. (c) Corresponding excess hysteresis area $A-A_0$ versus $R$.  
	}
\end{figure}

\begin{figure}[htbp]
	\includegraphics[width=\columnwidth]{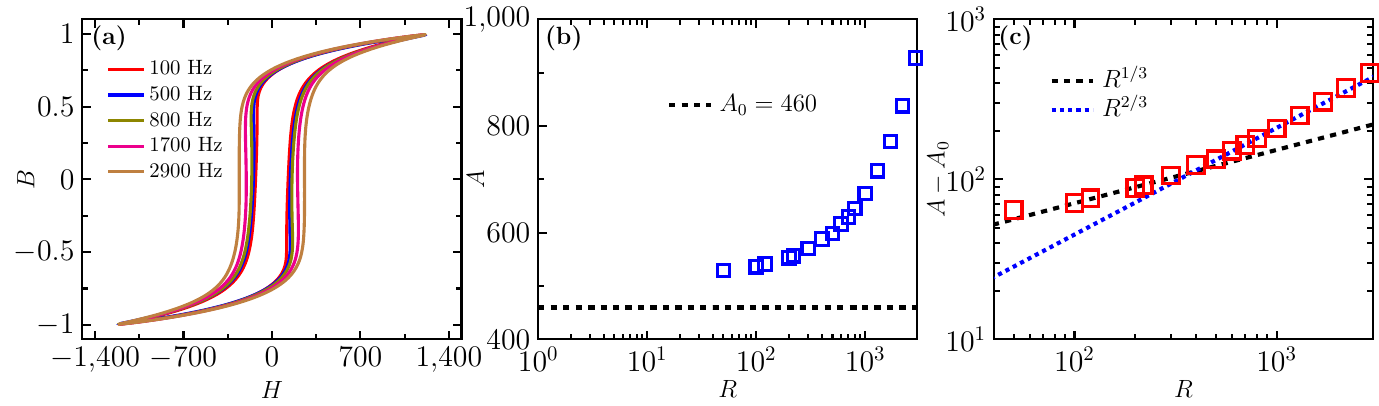}
	\caption{\label{SIfig:FeCoV} 
       {\bf Experimental results for the FeCoV alloy.}
    (a) $B$-$H$ curves measured in experiments of FeCoV alloy. (b) $A$ as a function of $R$  with $A_0=460$. (c) Corresponding excess hysteresis area $A-A_0$ versus $R$.  
	}
\end{figure}

\begin{figure}[htbp]
	\includegraphics[width=\columnwidth]{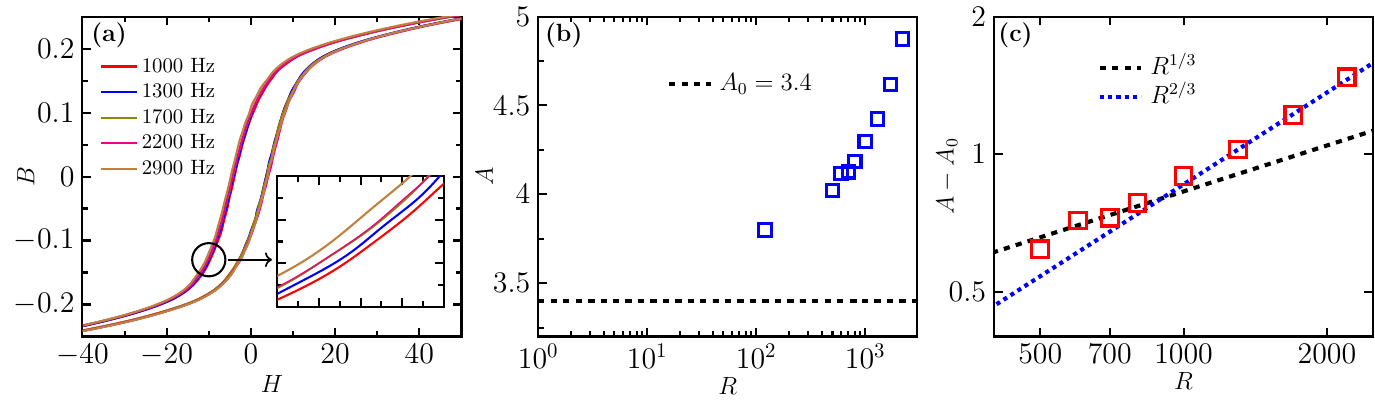}
	\caption{\label{SIfig:MnZnFe} 
     {\bf Experimental results for the MnZn ferrite.}
    (a) $B$-$H$ curves measured in experiments of MnZn Ferrite. Inset: enlarged $B-H$ curves. (b) $A$ as a function of $R$ with $A_0=3.4$. (c) Corresponding excess hysteresis area $A-A_0$ versus $R$.
	}
\end{figure}

\begin{figure}[htbp]
	\includegraphics[width=\columnwidth]{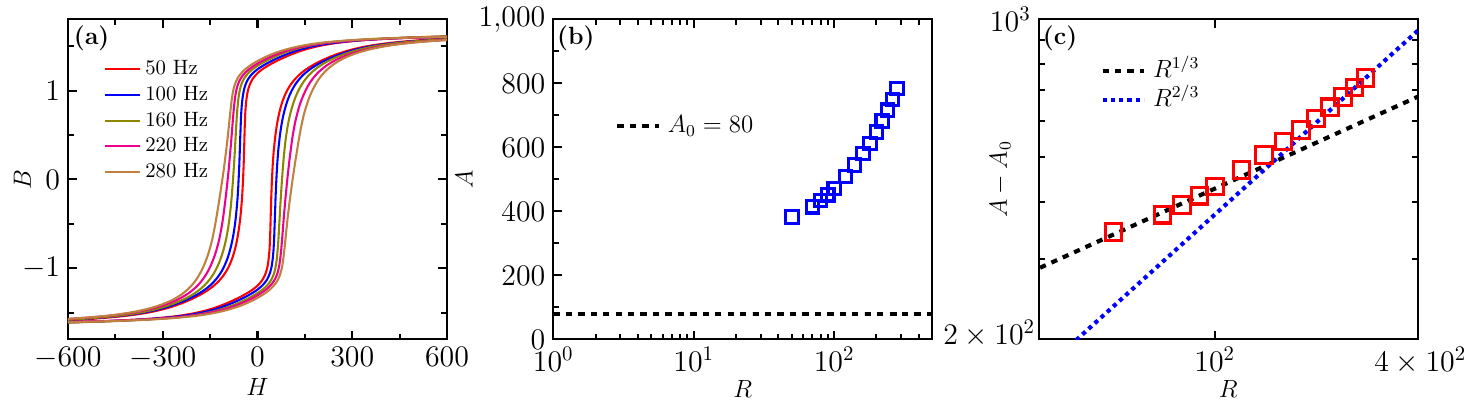}
	\caption{\label{SIfig:Si} 
     {\bf Experimental results for the Fe-3wt\%Si alloy.}
    (a) $B$-$H$ curves measured in experiments of Fe-3wt\%Si. (b) $A$ as a function of $R$ with $A_0=80$. (c) Corresponding excess hysteresis area $A-A_0$ versus $R$.
	}
\end{figure}

Detailed data are presented in Figs.~\ref{SIfig:nanocrystal}-\ref{SIfig:Si}
for the five magnetic materials, nanocrystalline alloy, permalloy, FeCoV alloy, MnZn ferrite and Fe-3wt\%Si alloy studied in our experiments.
The quasi-static area $A_0$ is determined using the strategy described in Sec.~\ref{SI:A_0}. 



\clearpage

\section{Additional results for Ising models}\label{SI:Ising}


\subsection{Determination of mean-field and non-mean-field regimes in the generalized Ising model}
For the generalized Ising model, we define the coordination number $z\equiv l^2-1$, which denotes the number of interacting neighbors for any central spin. The interaction range is controlled by $l$ (or $z$). We expect that this model is mean-field (MF) when $l$ is sufficiently large, while non-MF effects should emerge for small $l$. To examine such effects, we analyze the behavior of the critical temperature $T_c$ and the characteristic time $t_{\rm F}$ for global spin slip. 

\begin{figure}[htbp]
	\includegraphics[width=\columnwidth]
    {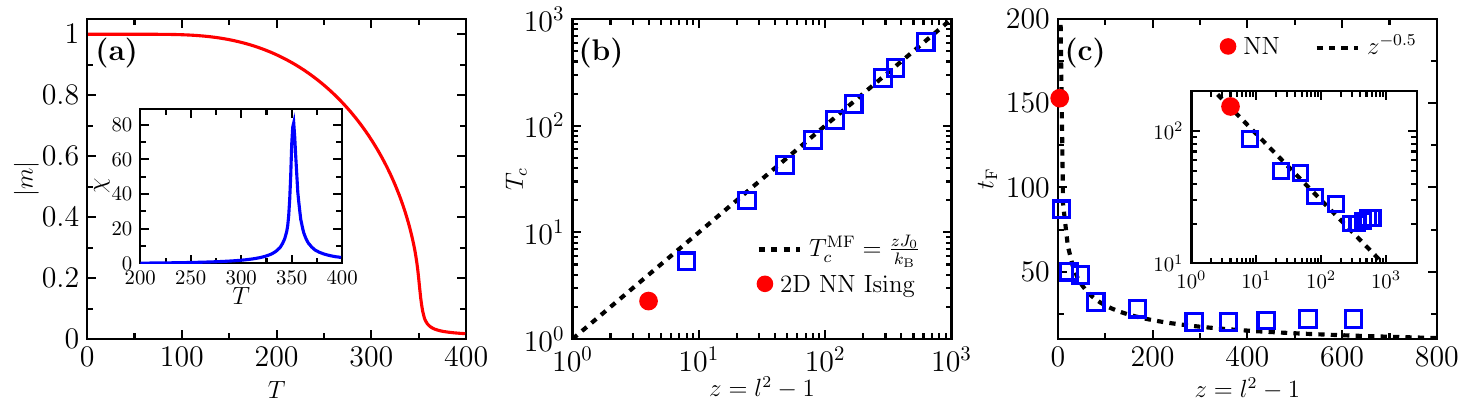}
    \caption{\label{SIfig:IsingTc}{\bf Non-MF effects on the critical temperature $T_c$ and global spin flip time $t_{\rm F}$.}
    (a) The absolute value of magnetization $m$ versus $T$ for the Ising model. Here $N=140$ and $l=19$. Inset: The corresponding susceptibility $\chi$ versus $T$. We can find that the critical temperature $T_c$ is about 349, defined by the maximum of the $\chi(T)$ curve. (b) $T_c$  versus the coordination number $z=l^2-1$. The dashed line denotes the MF relation,  $T_c^{\mathrm{MF}}=zJ_0/k_{\mathrm{B}}$. 
    The red solid circle denotes the analytical $T_c$ of the 2D nearest-neighbor (NN) Ising model.
    (c) The flip time $t_{\mathrm{F}}$ versus $z$. A line of  $t_{\mathrm{F}} = 300z^{-0.5}$ is drawn for reference. Inset: A log-log plot of the data. The red solid circle represents the simulation result for the NN Ising model.
    }
\end{figure}

In MF Ising models,   the critical temperature $T_c^{\mathrm{MF}}$ is proportional to $z$, $T_c^{\mathrm{MF}}=zJ_0/k_{\mathrm{B}}$, where $J_0$ and $k_{\mathrm{B}}$ are set to unity. It is found that the critical temperature $T_c$, which is numerically obtained from Fig.~\ref{SIfig:IsingTc}(a), satisfies the MF relation, $T_c^{\mathrm{MF}}=zJ_0/k_{\mathrm{B}}$, when $z> z^* \approx 100$ (see Fig.~\ref{SIfig:IsingTc}(b)).  Here the crossover coordination number $z^* \approx 100$ separates MF and non-MF regimes. 

The flip time $t_{\mathrm{F}}$ is measured by the time for all spins to flip from  down to  up in the fixed $T/T_c=0.5$ and $H=0.6$. Fig.~\ref{SIfig:IsingTc}(c) shows the change of $t_{\mathrm{F}}$ with $z$, from which we can see that $t_{\mathrm{F}}$ firstly decreases with $z$ and then remains unchanged. The region that $z$ keeps constant corresponds to the MF regime and the crossover point is at $z^* \approx 100$, which is consistent with the result in Fig.~\ref{SIfig:IsingTc}(b).

\begin{figure}[htbp]
	\includegraphics[width=0.6\columnwidth]{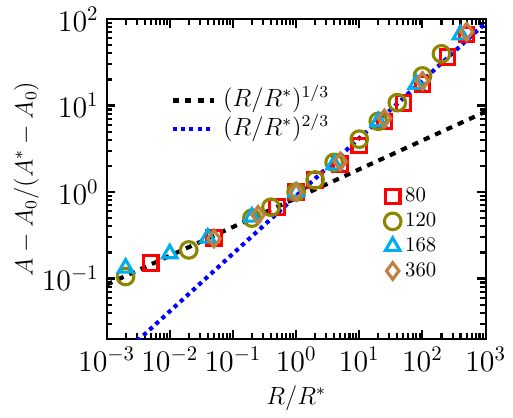}
	\caption{\label{SIfig:Isingz} 
    {\bf $(A-A_0)/(A^*-A_0)$ as a function of $R/R^*$ for Ising models with different $z$ (see the legend).
    } 
	}
\end{figure}

Based on the above analyses, we define the generalized Ising model with $z>z^* \approx 100$ as the MF Ising model. 
Figure~\ref{SIfig:Isingz} shows that the MF Ising models with different $z$ all satisfy the scaling Eq.~(\ref{eq:scaling}).


\subsection{Examples of hysteresis curves and scaling of hysteresis dispersion}
\begin{figure}[htbp]
	\includegraphics[width=\columnwidth]{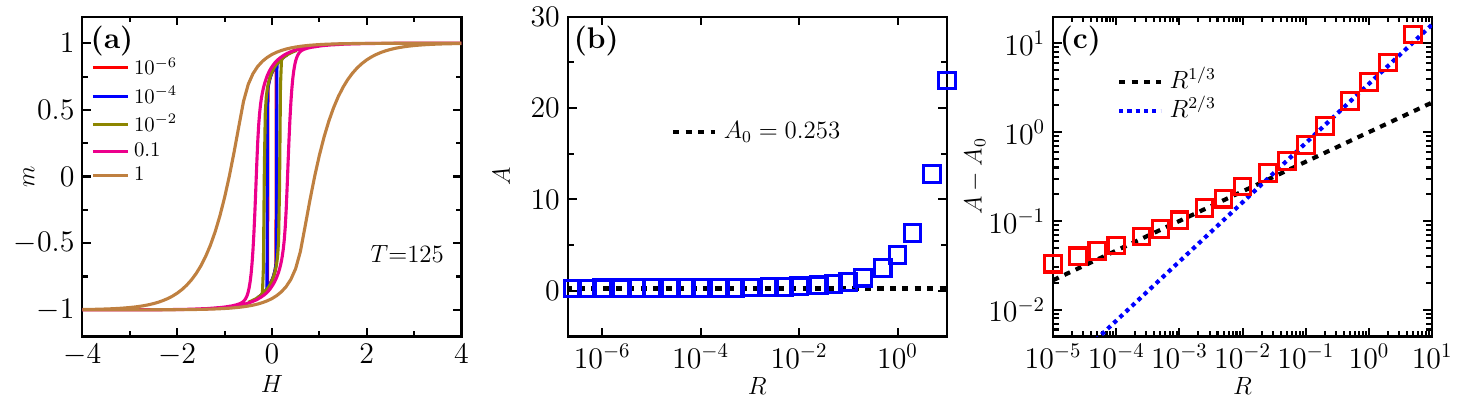}
	\caption{\label{SIfig:Ising}
    {\bf Simulation results for the MF Ising model with $l=13$.}
    (a) Magnetization - magnetic field ($m$-$H$) curves measured in simulations of the generalized Ising model, with a few different $R$ as indicated in the legend. (b) $A$ as a function of $R$.
    (c) Corresponding hysteresis area $A-A_0$ versus $R$.
	}
\end{figure}

\begin{figure}[htbp]
	\includegraphics[width=\columnwidth]{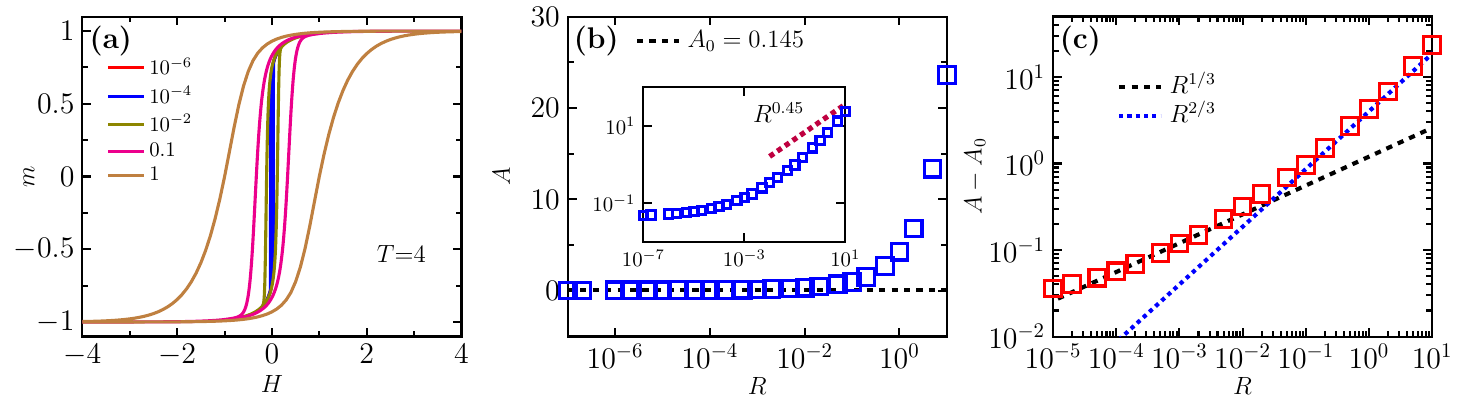}
	\caption{\label{SIfig:Ising3D}
    {\bf Simulation results for the 3D Ising model with the NN interaction.}
    (a) $m$-$H$ curves measured for the 3D NN Ising model, with a few different $R$ as indicated in the legend. (b) $A$ versus $R$ as a dynamic scaling.
    Inset: Log-log plot of $A(R)$, compared to $A \propto R^{0.45}$~\cite{PhysRevB.52.6550}.
    (c) Corresponding hysteresis area $A-A_0$ versus $R$. 
	}
\end{figure}

The simulation results of hysteresis curves for the MF Ising model with $l=13$ are presented in Fig.~\ref{SIfig:Ising}.  The simulation data of the 3D Ising model with the NN interaction are presented in Fig.~\ref{SIfig:Ising3D}, with  $A_0=0.145$.
In Ref.~\cite{PhysRevB.52.6550}, $A_0$ is considered to be 0 and the dynamic scaling is $A\propto R^{0.45}$. 
However, we find clear deviation from $A\propto R^{0.45}$ (see Fig.~\ref{SIfig:Ising3D} (b)-inset). Instead, setting $A_0=0.145$ gives a consistent result with Eq.~(\ref{eq:scaling}).



\subsection{Examination of the finite-size effect}
\begin{figure}[htbp]
	\includegraphics[width=\columnwidth]{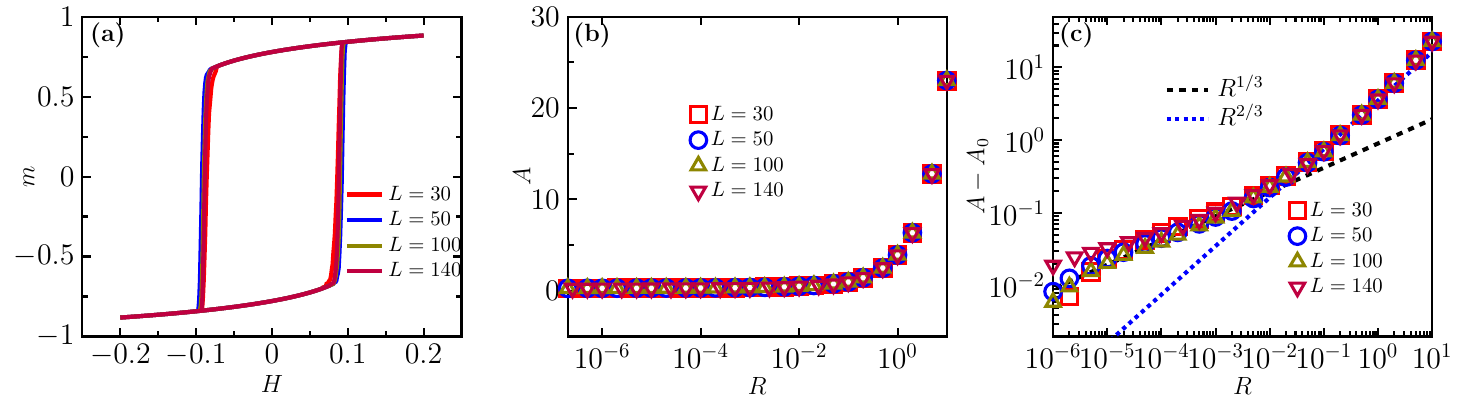}
	\caption{\label{SIfig:Isingsize}
    {\bf Size dependence of hysteresis for the MF Ising model with $l=13$.}
    (a) $m$-$H$ curves for different sizes of $L$, with the fixed $R=10^{-6}$. (b) $A$ versus $R$. (c) Corresponding dynamic scaling $A-A_0$ versus $R$.
	}
\end{figure}

\begin{figure}[htbp]
	\includegraphics[width=\columnwidth]{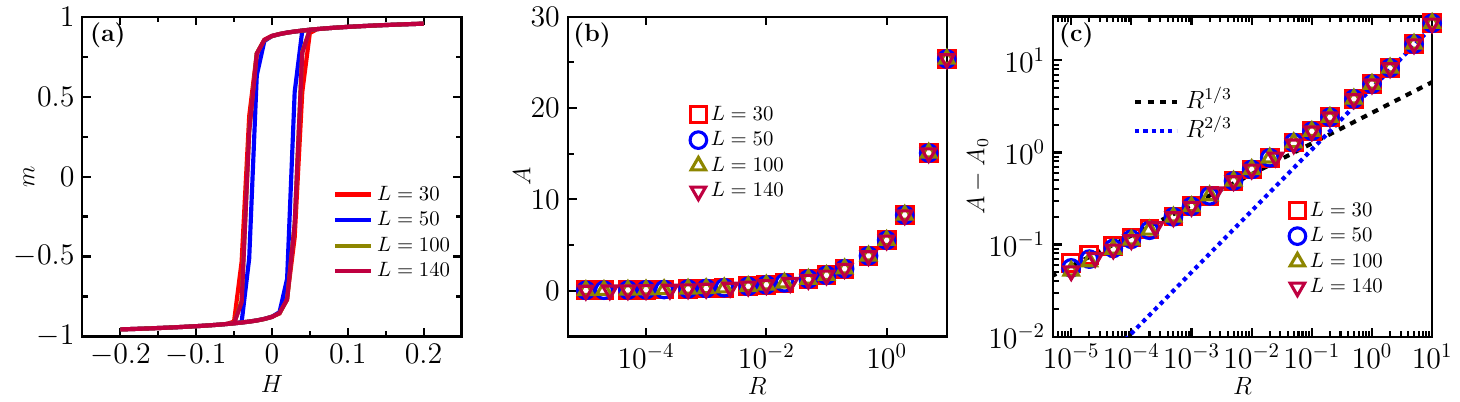}
	\caption{\label{SIfig:Isingsize2D}
    {\bf Size dependence of hysteresis for the 2D NN Ising model.}
    (a) $m$-$H$ curves for different sizes of $L$, with the fixed $R=10^{-4}$. (b) $A$ versus $R$. (c) Corresponding dynamic scaling $A-A_0$ versus $R$.
	}
\end{figure}

We examine the finite-size effect for the MF Ising model with $l=13$ at $T=125$ (Fig.~\ref{SIfig:Isingsize}) and for the 2D NN Ising model at $T=2.08$ (Fig.~\ref{SIfig:Isingsize2D}).  
The hysteresis curves and the corresponding dynamic scaling between $A-A_0$ and $R$ are found to be independent of the size $L$, consistent with the conventional wisdom that the finite-size effect is generally negligible for first-order phase transitions.  
This independence stands in sharp contrast to the strong finite-size effects near a critical point, where long-range spatial correlations are cut off by the finite system size. It suggests that the scaling in Eq.~(\ref{eq:scaling}) does not originate from long-range correlations or critical fluctuations. Moreover, since spatial correlations are not crucial, this also explains our observation that MF and non-MF systems display similar scaling behavior.


\clearpage
\section{Additional results for the MOF model}\label{SI:MOF}
\begin{figure}[htbp]
	\includegraphics[width=\columnwidth]{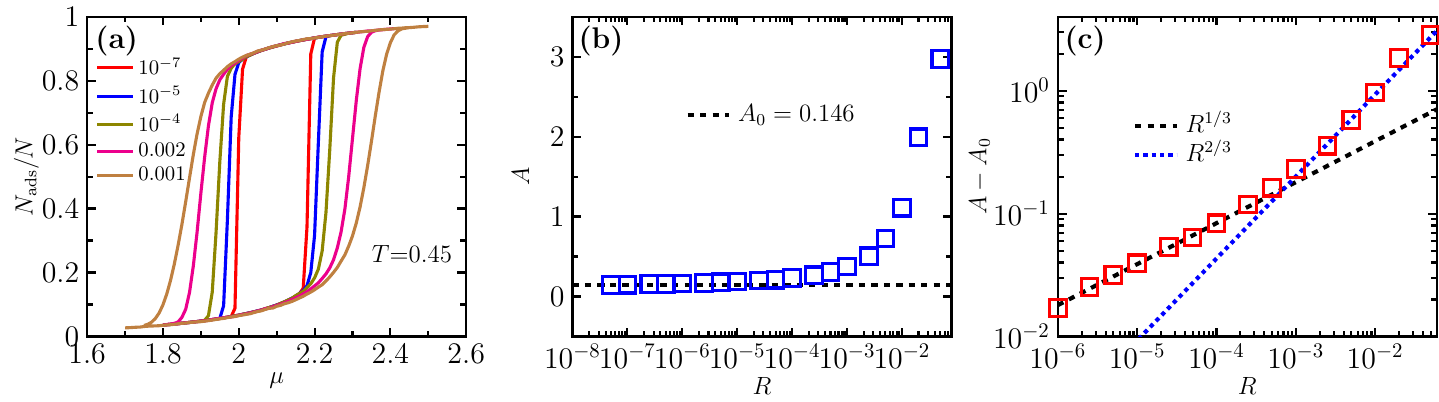}
	\caption{\label{SIfig:MOF} 
        {\bf Simulation results for the MOF model with $k=5$ and $a=0.6$.}
    (a) Fraction of guest particles adsorbed in the host MOF, $N_{\rm ads}/N$, as a function of the chemical potential $\mu$, for a few different $R$ (see the legend). (b) $A$ as a function of $R$, with $A_0=0.146$. (c) Corresponding hysteresis area $A-A_0$ versus $R$.
	}
\end{figure}

For the MOF model, we use two sets of parameters, $k=5$, $a=0.6$ and $k=3$, $a=2/3$, with critical temperatures  $T_c = 0.7$ and 0.53, respectively.
As an explicit example, the detailed results for the MOF model with $k=5$, $a=0.6$, and a fixed temperature $T=0.45$ are shown in Fig.~\ref{SIfig:MOF}.

\maketitle

\clearpage

\section{Determination of the quasi-static hysteresis area  $A_0$}\label{SI:A_0}

\begin{figure}[htbp]
	\includegraphics[width=0.7\columnwidth]
    {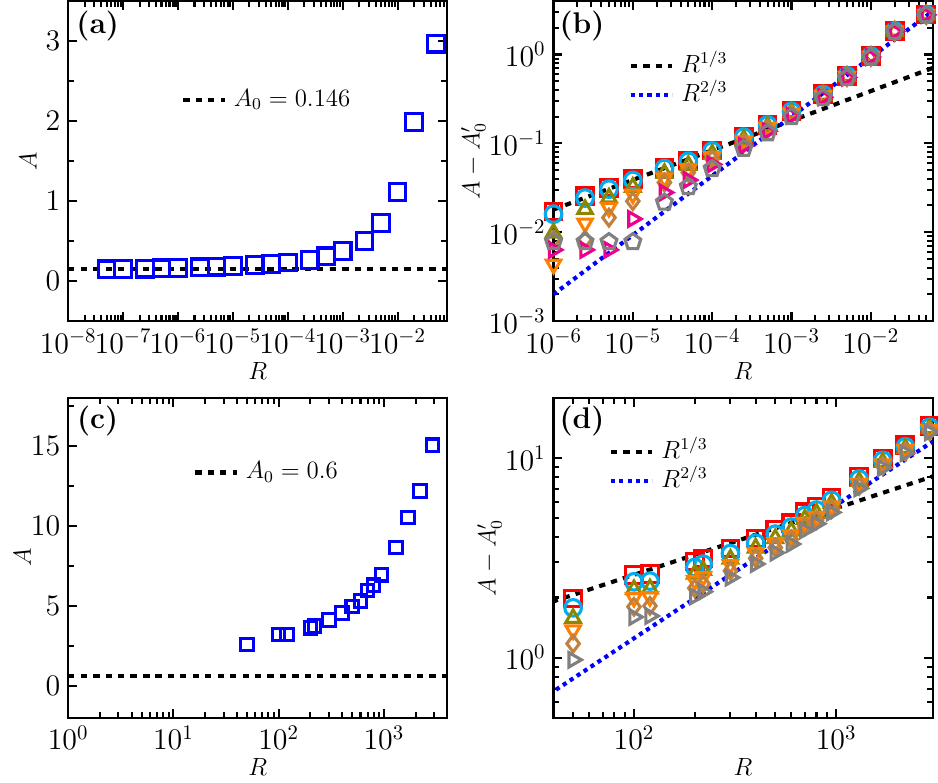}
	\caption{\label{SIfig:A0}
    {\bf Determination of $A_0$ in simulations and experiments.} (a) $A$ as a function of $R$ in simulations of the MOF model; $A_0$ is indicated by the horizontal line. 
    (b) $A - A_0'$ as a function of $R$ using different $A_0'$, for the simulation data of the MOF model, with $A_0'=$ 0.146, 0.148, 0.154, 0.160, 0.164, 0.172, 0.179, and 0.186 from top to bottom. (c) $A$ as a function of $R$ in experiments of the nanocrystalline alloy. (d) $A - A_0'$ as a function of $R$ using different $A_0'$, for the experimental data of  the nanocrystalline alloy, with $A_0'=$ 0.6, 0.8, 1.0, 1.2, 1.4, 1.6, 1.8, and 2.0 from top to bottom.
	}
\end{figure}

\begin{figure}[htbp]
	\includegraphics[width=0.5\columnwidth]
    {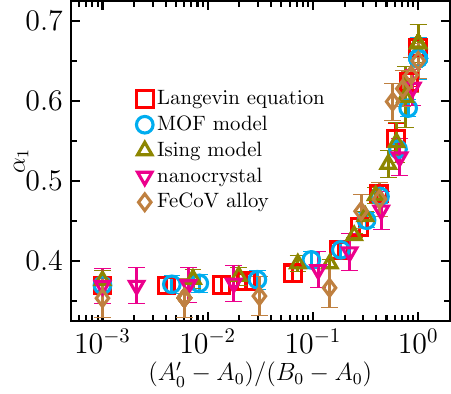}
    \caption{\label{SIfig:alphavsA0}
    {\bf Dependence of the small-$R$ exponent $\alpha_1$ on $A'_0$.}
    The exponent $\alpha_1$ is obtained from fitting the $A(R)$ data to $A-A'_0 \sim R^{\alpha_1}$,  for $R<R^*$, with the given $A'_0$ (see Fig.~\ref{SIfig:A0}). To collapse the data for different systems, $A'_0$ is shifted and rescaled as 
    $(A'_0 - A_0)/(B_0 - A_0)$, where $A_0$ is the value of $A'_0$ when $\alpha_1 = 1/3$, while   $B_0$ is the value of $A'_0$ when $\alpha_1 = 2/3$. 
	}
\end{figure}

For the given data of $A(R)$, by definition $A_0  = A(R \to 0)$. In our simulations, this quasi-static limit is accessible. Thus $A_0$ can be estimated by the plateau value of the $A(R)$ data in the small-$R$ regime (see Fig.~\ref{SIfig:A0}(a)). 
We have also examined the dependence of the exponent $\alpha$  on the value of $A'_0$. As shown in Fig.~\ref{SIfig:A0}(b), a series of $A'_0$ are used in the scaling plot $A-A'_0$ versus $R$. We find  that the exponent (i.e., $\alpha_1$) depends on $A'_0$ only  for $R< R^*$. This dependence is demonstrated in Fig.~\ref{SIfig:alphavsA0}: the exponent $\alpha_1$, obtained from fitting the data for $R<R^*$, decreases with a smaller $A_0'$ and approaches a constant $\alpha_1 \approx 1/3$.

In our experiments, the quasi-static regime cannot be accessed. As shown in Fig.~\ref{SIfig:A0}(c), the quasi-static  plateau of $A(R)$ is not strictly reached.  In this case, $A_0$ is obtained as an adjustable parameter. As shown in Fig.~\ref{SIfig:A0}(d), we adjust $A'_0$ and choose $A_0 = A'_0$ such that the small-$R$ exponent $\alpha_1$ becomes the desired value $1/3$. 
Increasing $A'_0$ from the minimum value $A_0$, the small-$R$ exponent $\alpha_1$ increases, reaching the athermal value $2/3$ at $B_0 = A'_0$. Thus, with the parameter $A'_0$ increasing from the minimum value $A_0$ to the maximum value $B_0$, the measured small-$R$ exponent $\alpha_1$ increases from $1/3$ to $2/3$ (Fig.~\ref{SIfig:alphavsA0}).
This approach gives an estimate of the quasi-static value $A_0$ when this limit cannot be accessed. The data in Fig.~\ref{SIfig:alphavsA0} shows that, the rescaled data of $\alpha_1$ versus $(A'_0 - A_0)/(B_0 - A_0)$ collapse for different systems, suggesting that the determined parameter $A_0$ gives a robust estimate of the small-$R$ exponent $\alpha_1$.


The above method gives a systematic approach to estimate $A_0$ in different systems. It has been noticed previously that $A_0$ is  system-dependent. 
In the athermal limit ($T = 0$), $A_0$ is determined by the stationary solution $H = a_2 \phi + a_4 \phi^3$ of the Langevin equation (see Fig.~\ref{fig:LEsigma}(a)). 
In this case,  $A_0$ is non-zero and the corresponding  exponent $\alpha_0 = 2/3$.
With thermal noises, it is found in many works that $A_0 \to 0$ when the rate $R$ is small, including  the 2D nearest-neighbor Ising model~\cite{PhysRevA.42.7471}, Langevin dynamics of $(\Phi^2)^2$~\cite{PhysRevB.42.856} and $(\Phi^2)^3$ models~\cite{PhysRevB.43.3373}, with a corresponding  exponent $\alpha_1\approx1/3$.

\clearpage
\section{Derivation of the scaling function Eq.~(6)}\label{SI:derivation}

\subsection{Perturbation theory for the Langevin equation}
\label{SI:perturbation}
To derive Eq.~(\ref{eq:CorrectScaling21}) in the main text, we generalize the perturbation theory~\cite{krapivsky2010kinetic,holmes2012introduction}. This theory is initially developed to derive the scaling exponents  for the athermal  Langevin equation ($\xi = 0$)~\cite{PhysRevE.108.024101}. Here we extend the method to higher-order perturbation terms to include the effects of thermal noise ($\xi > 0$).

Consider the Langevin equation Eq.~(\ref{eq:LE2}) in the main text, 
\begin{align}
	\frac{d \phi}{d t} = -(a_2\phi + a_4\phi^3 - H) + \xi,
\end{align}
where $\lambda $ has been set to unity. Let $H=R t$, 
we  obtain,
\begin{align}\label{SIeq:transLE}
	R\frac{d \phi}{d H} = -a_2\phi - a_4\phi^3 + H + \xi.
\end{align}

Considering the rate $R$ as a small parameter near the quasi-static limit  $R\rightarrow0$, we can expand $\phi$  with $R$ as,
\begin{equation}
\phi = R^0\phi_0 + R^1\phi_1 + R^2\phi_2 + ...
\end{equation}
Matching terms at different orders then gives the scaling exponents.

\subsubsection{Athermal case}
When $\xi = 0$ ($T=0$), the stationary solution ($d \phi/d H =0$) of Eq.~(\ref{SIeq:transLE}) is given by, to the lowest order $R^0$, 
\begin{equation}
a_2\phi_0+a_4\phi_0^3=H.
\end{equation}
This equation gives the spinodal points, 
($H_s$, $\phi_s$)=($\frac{2a_2}{3}\sqrt{\frac{-a_2}{3a_4}}$, $\sqrt{\frac{-a_2}{3a_4}}$).
Shifting the variables by the spinodal values, one proposes the following ansatz,

\begin{align}\label{SIeq:cornervariable_ath}
\nonumber	H & = H_s +  R^{\alpha} \tilde{H}_1  + \ldots,\\ 
        \phi & = \phi_s + R^{\beta}\tilde{\phi}_1 + R^{2\beta}\tilde{\phi}_2 + \ldots,
\end{align}
where the two exponents $\alpha$ and $\beta$ will be determined by scaling matching. 

Substituting Eq.~(\ref{SIeq:cornervariable_ath}) into Eq.~(\ref{SIeq:transLE}) and setting $\xi = 0$, we get the first-order equation, 
\begin{align}\label{SIeq:simexpansion1}
	R^{1+\beta-\alpha}\frac{d \tilde{\phi}_1}{d \tilde{H}_1} = - 3a_4R^{2\beta}\phi_s\tilde{\phi}_1^2 +\tilde{H}_1R^{\alpha}.
\end{align}
Because the powers of $R$ in different terms should match, we have
\begin{align}
	1 + \beta - \alpha = 2\beta = \alpha,
\end{align}
and therefore, $\beta=\frac{1}{3}$ and $\alpha=\frac2 3$. This derivation gives the athermal exponent (large-$R$) exponent $\alpha_0 = 2/3$ in Eq.~(\ref{eq:scaling}).

\subsubsection{Thermal case}
When $T > 0$, the zero-th order stationary equation of the Langevin dynamics is
\begin{equation}
a_2\phi_0+a_4\phi_0^3=H + \xi.
\end{equation}
The solution of $dH/d\phi_0 = 0$ gives the effective spinodal point  ($H_0(T)$, $\phi_0(T)$), whose explicit expression is unknown. In the athermal limit, 
($H_0(T \to 0), \phi_0(T \to 0)$) = ($H_s$, $\phi_s$)=($\frac{2a_2}{3}\sqrt{\frac{-a_2}{3a_4}}$, $\sqrt{\frac{-a_2}{3a_4}}$).

The expansion of ($H,\phi$) around ($H_0, \phi_0$) is assumed to have the following ansatz,
\begin{align} \label{SIeq:cornervariable}
    \nonumber&H = H_0 + R^\alpha \tilde{H}_1 +  R^{\alpha + \delta} \tilde{H}_2 + \ldots\\
         &\phi = \phi_0 + R^{\beta}\tilde{\phi}_1 + R^{2\beta}\tilde{\phi}_2 + \ldots 
\end{align}
By substituting Eq.~(\ref{SIeq:cornervariable}) into Eq.~(\ref{SIeq:transLE}), we get
\begin{align}\label{SIeq:expansion}
    R^{1+\beta-\alpha}\frac{d \tilde{\phi}_1}{d \tilde{H}_1} 
    = -a_2R^{\beta}\tilde{\phi}_1-3a_4R^{\beta}\phi_0^2\tilde{\phi}_1 - 3a_4R^{2\beta}\phi_0\tilde{\phi}_1^2  
    + R^{\alpha}\tilde{H}_1 +R^{\alpha+\delta} \tilde{H}_2.
\end{align}
To the first-order, $\tilde{\phi}_1$ satisfies a similar equation as Eq.~(\ref{SIeq:simexpansion1}),
\begin{align}
	R^{1+\beta-\alpha}\frac{d \tilde{\phi}_1}{d \tilde{H}_1} = - 3a_4R^{2\beta}\phi_0\tilde{\phi}_1^2 +\tilde{H}_1 R^{\alpha}.
\end{align}
The remaining terms in Eq.~(\ref{SIeq:expansion}) are,
\begin{align}\label{SIeq:expansion2}
	0 = -a_2R^{\beta}\tilde{\phi}_1-3a_4R^{\beta}\phi_0^2\tilde{\phi}_1 + \tilde{H}_2 R^{\alpha+\delta}.
\end{align}
Match powers of $R$ again and note that $\alpha=\frac2 3$, $\beta=\frac1 3$, we have
\begin{align}
	\frac1 3 = \beta = \alpha+\delta,
\end{align}
and $\delta=-\frac1 3$. 

The above analysis shows that the  first line in Eq.~(\ref{SIeq:cornervariable}) can be expressed as
\begin{align}
	H-H_0\sim R^{2/3}[1+c(T)R^{-1/3}],
	\label{SIeq:CorrectScaling}
\end{align}
where we let $\tilde{H}_2(T) = c(T) \tilde{H}_1$ (note that $\tilde{H}_2$ depends on $T$).
Since the area of the hysteresis loop $A$ is proportional to $H$, it satisfies
\begin{align}
	A-A_0\sim R^{2/3}[1+c(T)R^{-1/3}],
	\label{SIeq:CorrectScaling2}
\end{align}
where $A_0$ denotes the hysteresis area of the quasi-static state. The small-$R$ exponent $\alpha_1=1/3$ (see Eq.~(\ref{eq:scaling}))) arises in Eq.~(\ref{SIeq:CorrectScaling2}).

\subsection{Scaling of $c(T)$}
\label{SI:CUS}
Next we consider $c(T)$ in Eqs.~(\ref{SIeq:CorrectScaling}) and (\ref{SIeq:CorrectScaling2}).
The complete universal scaling (CUS)~\cite{PhysRevLett.95.175701,ChinPhysLett.41.100502},
\begin{align}\label{SIeq:CUS}
	(\phi-\phi_s) R^{-\frac1 3} = \Phi[ a_2R^{-\frac1 3}, (H-H_s)R^{-\frac2 3}, T R^{-1}, a_4R^{\frac1 3}],
\end{align}
suggests that the scale invariant form for the temperature is, $TR^{-1}$, i.e., $c(T) = c_0 T^{1/3}$. 
Thus Eqs.~(\ref{SIeq:CorrectScaling}) and (\ref{SIeq:CorrectScaling2}) can be written as,
\begin{align} \label{SIeq:expression3}
		H-H_0\sim  R^{2/3}\left[1+c_0 \left(\frac{T}{R} \right)^{1/3}\right],
\end{align}
and 
\begin{align} \label{SIeq:CorrectScaling3}
		A-A_0\sim  R^{2/3}\left[1+c_0 \left(\frac{T}{R} \right)^{1/3}\right].
\end{align}

\begin{figure}[htbp]
	\includegraphics[width=0.7\columnwidth]{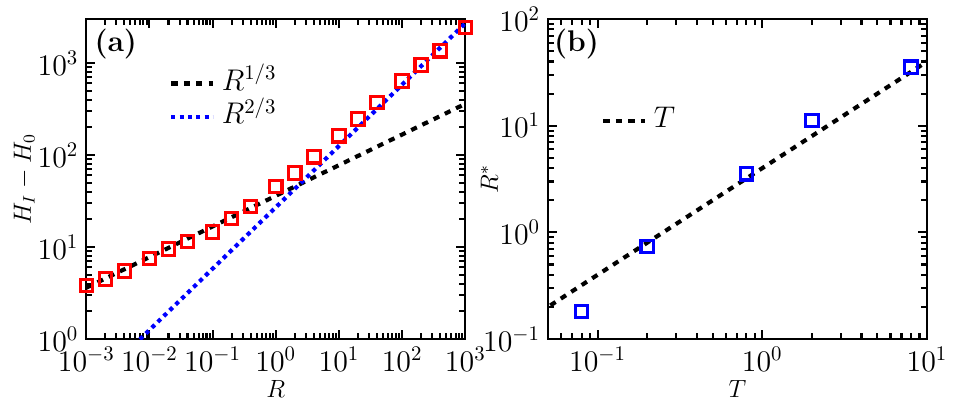}
	\caption{\label{SIfig:H_R}
    {\bf Validation of Eq.(\ref{SIeq:scaling}).}
    (a) $H_I-H_0$ versus $R$ at $T=0.2$. 
    (b) $R^*$ versus $T$.
    Data are obtained from the Langevin equation with $\lambda = 0.0005$.  The values of $H_0$ are estimated  at the jump point of the hysteresis curve corresponding to a very small rate, $R=0.0001$.
	}
\end{figure}

Equation~(\ref{SIeq:expression3}) gives a scaling of $H$, which is similar to Eq.~(\ref{eq:scaling}):
\begin{equation}
H - H_0 \propto 
\begin{cases} 
R^{1/3}, & R < R^*, \\ 
R^{2/3}, & R > R^*, 
\end{cases}
\label{SIeq:scaling}
\end{equation}
where $R^* \propto T$. This scaling is explicitly verified for the inflection point $(H_I, \phi_I)$, which is the location of the maximum slope $d\phi/dH$ of the hysteresis curve (see Fig.~\ref{SIfig:H_R}). 

\end{document}